\documentstyle[psfig,12pt]{report}
\input epsf.tex
\newcommand{\be}{\begin{equation}}
\newcommand{\ee}{\end{equation}}
\newcommand{\bea}{\begin{eqnarray}}
\newcommand{\eea}{\end{eqnarray}}

\def\path#1{PICT/#1}

\newcommand{\Det}{\mathop{\rm Det}}
\newcommand{\PD}{\partial}

\def\href#1#2{#2}
\newcommand{\req}[1]{~(\ref{#1})}
\newcommand{\gs}{\mbox{$g_s$}}            
\newcommand{\ap}{\mbox{$\alpha^\prime$}}  
\newcommand{\ls}{\mbox{$l_s$}}            
\newcommand{\g}[1]{g_{YM_{#1}}}
  
\begin{document}

\begin{titlepage}
\begin{center}
\vspace{2cm}
{\huge \bf Born-Infeld Action in String Theory }

\vspace{3cm}

{\Large Konstantin George Savvidy }

\vspace{5cm}

{A DISSERTATION \\
PRESENTED TO THE FACULTY \\
OF PRINCETON UNIVERSITY\\
IN CANDIDACY FOR THE DEGREE\\
OF DOCTOR OF PHILOSOPHY}

\vskip 1.5truein
{RECOMMENDED FOR ACCEPTANCE \\
BY THE DEPARTMENT OF PHYSICS}
\vskip .2truein
\centerline{JUNE 1999}
\vskip .5truein

\end{center}
\nonumber
\end{titlepage}

\newpage
\nonumber
\vspace{2in}
\copyright Copyright by Konstantin G Savvidy, 1999.~~~
All rights reserved.
\nonumber
\newpage
\addcontentsline{toc}{chapter}{Abstract}
\begin{center}
{\huge\bf Abstract}
\end{center}
\addtolength{\baselineskip}{+.2cm}
In this thesis I present applications of the Born-Infeld action to the 
description of D-branes, mostly in the context of type IIB string 
theory. I briefly review the relevant aspects of the general 
formalism, including the efforts toward generalization of this action to the 
non-Abelian case. In the second Chapter I present the construction
of a fundamental string attached to a threebrane, whereby the string
is made out of the wrapped brane. This interpretation withstands
a rather detailed scrutiny, including the magnitude of the 
static tension and of the B-I charge;
and
Polchinski's mixed Dirichlet/Neumann boundary conditions for small
perturbations of the string. The latter arise dynamically from the full
non-linear equations of motion. Also, the Born-Infeld charge is shown
to produce the correct form of dipole radiation under the influence of
harmonic oscillations of the string.

The third Chapter arises out of two papers on the application of
these ideas to the AdS/CFT correspondence. Using the essentially
non-perturbative nature of the B-I action we were able to construct
in the near-horizon geometry of N D3-branes
the color-singlet baryon vertex of the SU(N) gauge theory as a 
fivebrane wrapped on the $S^5$ sphere, with B-I strings playing
the role of the quarks. Also, in the asymptotically flat background
we obtain a smooth classical 
description of the Hanany-Witten phenomenon, whereby one drags
a flat fivebrane across the stack of threebranes, and B-I strings
stretching between the two
appear and disappear without any discontinuity.

In the case of non-extremal background we construct baryons of 
the three-dimensional non-supersymmetric theory, the low energy 
dynamics of which was conjectured to be equivalent to ordinary QCD.
Our approach allows to exhibit confining properties of color flux 
tubes, i.e. the dependence of the tension of the flux tube on its 
color content. We do this by building an extended baryon, where 
a fraction $\nu$ of the total of N quarks is removed to a finite distance.

\pagenumbering{roman}
\setcounter{page}{3}
\pagebreak
\addtolength{\baselineskip}{-.2cm}
\tableofcontents

\pagebreak

\addtolength{\baselineskip}{+.1cm}
\newpage
\chapter{Introduction and History}
\label{intro}
\pagenumbering{arabic}
\setcounter{page}{1}
\addtolength{\baselineskip}{+.2cm}
Research into various aspects of D-branes in the past few years
has become synonymous with the New String Theory. Beginning 
with the first work by Dai, Leigh and Polchinski in 1989 \cite{pol},
and the key discovery by Polchinski \cite{rr} of the fact that D-branes carry
RR charges led to an incredible amount of activity surrounding branes.
D-branes, in particular the D2-brane and the NS5-brane are presently 
thought to be the fundamental objects of M-theory, which is conjectured
to incorporate all the known species of string theory, plus the 
11-dimensional supergravity. An important milestone was the 
construction of the corresponding supergravity solutions by
G.~Horowitz and A.~Strominger \cite{stroho}.
This work itself built on the previous construction of the SUGRA 
solution for an infinite fundamental string by Dabholkar and Harvey 
\cite{daha}.

The Born-Infeld action, sometimes also referred to as Dirac-Born-Infeld \cite{bi,dirac}
action is the effective action for low-energy degrees of freedom of the
D-brane \cite{leigh}. Even before that, the consistency of the $\sigma$-model 
for the world-sheet of the string was shown to require 
the background field to be described by BI action \cite{FT},
just like in the general curved background requiring consistency of string theory
leads to the Einstein-Hilbert action \cite{abouelsaood}. It should come as no surprise then, 
that BI is also invariant under general coordinate transformations,
in fact it can be thought of as a theory with a more general metric, where the
anti-symmetric part of the metric tensor is identified with the electromagnetic field.
This was the thinking of Born and Infeld \cite{bi} when they presented the final
version of their modified electrodynamics. The original motivation was to
get a better behaved theory, one that would not have the Coulomb self-energy
divergence, etc... Also one might hope that the corresponding quantum
field theory might be better behaved too, but progress is of course 
hindered by the fact that the action contains an infinite number of 
unrenormalisable interactions.
 
As explained above, the key difference with 
previously known p-branes and supermembranes 
\cite{supermembrane,supermembrane1,supermembrane2} is the inclusion
of a worldvolume electromagnetic field. This field corresponds on the 
string theory side to virtual states of an open string, where both
ends are fixed to move in the worldvolume of the brane.

The starting point for this thesis is the paper by Callan and
Maldacena \cite{cm}, hereafter referred to as CM.
They, noting that most treatments of D-branes were done using
the linearized theory, undertook to fully account  for the
specific non-linearities of the Born-Infeld
{\it super-electrodynamics}. The static features of 
intersecting branes are determined by SUSY and BPS arguments,
however dynamical issues have to be explored using the full 
Born-Infeld action. They construct, first and foremost, the 
 solution for a fundamental string ending on a D-brane, in which
the junction-point of the string manifests itself as a point Coulomb charge
in the world-volume of the brane. They also find a non-BPS solution
corresponding to a brane and an anti-brane joined by e.g. a
fundamental string.

This thesis is based on a series of papers that arose directly
out of the approach Callan and Maldacena took in their paper.
For the sake of convenience I list these papers in the order in which they
are presented in this thesis:
\begin{itemize}
\item ``Neumann Boundary Conditions from Born-Infeld Dynamics,''\\
by K. Savvidy, G. Savvidy, {{\tt hep-th/9902197}}
\item ``Brane Death via Born-Infeld String'', by K. Savvidy, {{\tt hep-th/9810163}} 
\item ``Baryons and String Creation from the Fivebrane Worldvolume Action,''\\ 
by C.G. Callan, A. Guijosa, K. G. Savvidy, {{\tt hep-th/9810092}}
\item ``Baryons and Flux Tubes in Confining Gauge Theories from Brane Actions,''\\
by C.G. Callan, A. Guijosa, K. G. Savvidy, O. Tafjord, {{\tt hep-th/9902197}} 
\end{itemize}
I have used the texts of these papers to large extent verbatim,
especially in the case of the latter two papers, in this thesis.

In Chapter 1 I review some general facts about the Born-Infeld action.
In particular, section \ref{gen_cov} deals with the two possible representations
of the action, the manifestly generally covariant one, and the dimensionally
reduced version which is necessarily in the "static" gauge. Section 
\ref{nonabelian} discusses briefly the subject that hasn't received full 
development as yet, that of non-Abelian generalization of the B-I action.

In Chapter 2 I present some applications of ideas from CM to 
threebrane physics. Specifically, sections \ref{dirichlet} and \ref{neumann}
show how Polchinski's open string boundary conditions arise dynamically
from the B-I  treatment of the F-string attached to a D3-brane. 
In CM it was shown that
excitations which are normal to both the string and the 3-brane behave as if
they had Dirichlet boundary conditions at the point of attachement.
We show that certain excitations of the F-string/D3-brane 
system can be shown to obey   Neumann boundary conditions by considering 
the Born-Infeld dynamics of the F-string (viewed as a 3-brane cylindrically
wrapped on an $S_2$). In CM it was shown that
excitations which are normal to both the string and the 3-brane behave as if
they had Dirichlet boundary conditions at the point of attachment. 
In \cite{neumannP} 
we show that excitations which are coming down the string with a 
polarization along a direction parallel to the brane are almost completely
reflected just as in the case of all-normal excitations, but the end of
the string moves freely on the 3-brane, thus realizing Polchinski's
open string    Neumann
boundary condition dynamically. In the low energy limit 
$ \omega \rightarrow 0$, i.e. for wavelengths much larger than the string
scale only a small fraction $ \sim \omega^4$ of the energy escapes in
the form of dipole radiation.
The physical interpretation is that a string attached to the 3-brane 
manifests itself as an electric charge, and waves on the string cause
the end point of the string to freely oscillate and therefore produce 
electromagnetic dipole radiation in the asymptotic outer region.

Section \ref{brane_death} contains the calculation of the brane-anti-brane
potential, which can be used to capture some new features of the quasi-classical 
tunneling of the brane-anti-brane pair joined by an F-string. This is
of relevance to a subject that has since attracted certain attention,
the brane-anti-brane annihilation: see e.g. A.~Sen, "Tachyon condensation.." 
and E.~Witten "D-branes and K-theory". Our calculation shows that for 
a distance between the branes that is larger than some critical distance,
determined by the string coupling, the stretched string is indeed quasi-stable,
while for shorter distances it is tachyonic.

The attraction between
the branes causes them to approach and at some point to tunnel, 
because the action of the bounce solution goes to zero.
The energy of the solution at the top of the barrier, the sphaleron, goes like
$\sim D^3$ for large separations D, while the energy of the string is proportional
to its length D.

Chapter 3 has a slightly different flavor. There, the B-I action
is used to describe a fivebrane placed in the supergravity background
of N coincident D3-branes. In the near-horizon limit, according to Witten,
the background five-form RR field induces N units of electric charge
on the fivebrane, which should be interpreted as N fundamental strings
attached to it. These strings play the role of quarks in the AdS/CFT correspondence,
and the configuration as a whole is interpreted as a color neutral baryon
of SU(N) {\it super}-YM theory.

In essence, B-I is known to incorporate corrections to string theory to all
orders in $\ap$, and thus allows us to construct an essentially non-perturbative
object on the gauge theory side: the baryon. In addition, this approach is
self consistent, since B-I action is only valid for slowly varying fields
(the action does not contain derivative-dependent corrections to the effective string
action). Our solutions  have fields that are everywhere smooth thus
ensuring the consistency of our description.

The extension of these to the full D3 metric results in a nice 
classical description of the Hanany-Witten phenomenon, whereby strings are
created when a flat D5-brane is pulled through the threebrane stack.
A surprising byproduct is the configuration where N/2 strings are stretched 
to the threebranes on either side, and change their orientation when one
crosses the threebrane. Thus the total change in the number of strings
is N, as expected. This configuration%
\footnote{It was found for the case of 0-brane in \cite{dfk}.},
is characterized by zero force on the 5-brane, as the attraction due to the
tension of the string is exactly balanced by electrostatic repulsion. 
These considerations
seem to suggest the actual existence of the 1/2 string state, the proper
interpretation of which is at present not clear.

In the section \ref{3dsec} we use near-horizon geometry
of non-extremal D3-branes to find embeddings of a D5-brane which describe
baryonic states in three-dimensional QCD.
In particular, we
construct solutions corresponding to a baryon made of $N$ quarks,
and study what happens when some fraction $\nu$ of the total
number of quarks are bodily moved to a large spatial separation
from the others. The individual clumps of quarks are represented
by Born-Infeld string tubes obtained from a D5-brane whose
spatial section has topology ${\bf R}\times {\bf S}^4$.
They are connected by a confining color flux tube, described
by a portion of the fivebrane that runs very close and parallel
to the horizon. We find that this flux tube has a tension with a
nontrivial $\nu$-dependence (not previously obtained by other methods).

We also discuss related questions including binding energies,
string creation, and the appearance of new energy scales.


\section{General Covariance of the Born-Infeld Action}
\label{gen_cov}
This section closely follows \cite{schwarz} by 
M. Aganagic, C. Popescu and J.H. Schwarz and \cite{gibbons}
by G.W. Gibbons in the treatment of general covariance of the
Born-Infeld action. Born and Infeld \cite{bi} realized the final version
of their  non-linear electrodynamics through a manifestly 
covariant action. In modern language this can be expressed by
saying that the world-volume theory of the brane is described 
by the action
\be
S_{(p)}=-\frac{1}{(2\pi)^pg_s}\int d^{p + 1} \sigma \sqrt{- {\rm {\rm det}} (G_{\mu\nu} + {F}_{\mu\nu})}
\ee
where {\cal F} is the world-volume electromagnetic field strength,
measured in units in which $2\pi\ap =1$.
G is the induced metric on the brane
\begin{equation}
G_{\mu\nu} = \eta_{mn} \partial_\mu X^m \partial_\nu X^n.
\end{equation}

The action is invariant
under arbitrary diffeomorphysms of the world-volume. One way
of fixing this freedom is to adopt the so-called ``static gauge''
for which the world-volume coordinates  are equated with the first
$p+1$ space-time coordinates:
\be
X^{\mu}  \equiv \sigma^{\mu},~~ \mu=0,1,..,p~.
\ee
This ``static gauge'' description is most convenient if the brane is indeed 
positioned along those directions. 

The rest of the coordinates become world-volume fields
\be
X^{m} \equiv  \phi^{m},~~ m=p+1,..,9~.
\nonumber
\ee
The B-I action becomes
\begin{equation}
S'_{(p)}=-\frac{1}{(2\pi)^pg_s}\int d^{p+1} \sigma \sqrt{-\det (\eta_{\mu\nu} + \partial_\mu \phi^i \partial_\nu
\phi^i + F_{\mu\nu})}~ .
\end{equation}
Note that this is in some sense a modification of pure B-I: it
has extra scalar fields $\phi$.

Another way of obtaining the same is to dimensionally reduce
from 10D flat-space B-I by assuming that fields depend only
on the first $p+1$ coordinates , where $p$ is the number of
space dimensions of the brane. In this picture the extra
components  of the gauge field , $A_{p+1}...A_9$ come to 
represent the massless transverse excitations $\phi_{p+1}...\phi_9$.
In fact, they are the Goldstone bosons associated with spontaneously
broken translational symmetries. 
In this way one gets to compute a $10 \times 10$ determinant 
instead of a $4 \times 4$ determinant. The general coordinate
invariance is not apparent in this approach, and in fact it is hard
or impossible to use if a more general parametrization of the
brane is desired, or the brane is immersed 
into an already curved background. A simple illustration
of the equivalence of this two forms of the action
is to consider an elecrostatic case, with only one scalar
field excited. The corresponding determinants are%
\footnote{We specialize to the case of three-brane here.} 
\be
\left(
\mbox{
$\begin{array}{ccccc}
-1  & E_1 & E_2 & E_3 & 0       \\
-E_1& 1   & 0   & 0   &\PD_1\phi\\
-E_2& 0   & 1   & 0   &\PD_2\phi\\
-E_3& 0   & 0   & 1   &\PD_3\phi\\
0   & -\PD_1\phi & -\PD_2\phi & -\PD_3\phi &1
\end{array}$}
\right) 
vs.
\left(
\mbox{$\begin{array}{cccc}
-1  & E_1                   & E_2                    & E_3       \\
-E_1& 1 + \PD_1\phi^2       &  \PD_1\phi \PD_2\phi   & \PD_1\phi \PD_3\phi  \\
-E_2& \PD_1\phi \PD_2\phi   & 1+\PD_2\phi^2          &\PD_2\phi \PD_3\phi\\
-E_3& \PD_1\phi \PD_3\phi   & \PD_2\phi \PD_3\phi    & 1+\PD_3\phi^2 
\end{array}$}
\right) 
\ee
It is easy to see that the two are equivalent by simple 
row and column subtractions.

The discussion so far, and in most applications, concerns 
the ``bosonic'' action. However it is important to remember
that it is  a SUSY theory that we are dealing with. Some
non-trivial issues are presented in the paper \cite{schwarz}
by Aganagic, Popescu and Schwarz. 
In the first,
manifestly general-covariant representation, the supersymmetric
theory consists of $X^m$ , the flat 10-dimensional space coordinates;
the $U(1)$ gauge field $A_{\mu}$ on the world-volume; and a pair
of  Majorana-Weyl spinors $\theta_1$ and $\theta_2$. 
These spinors are of the same or opposite chirality depending on 
whether we consider type IIA or IIB theory (even/odd p).
The resulting theory possesses a high degree of symmetry:
\begin{itemize}
\item global super-Poincare group in 10D
\item local general coordinate invariance
\item local fermionic ``kappa'' symmetry
\item local U(1) gauge invariance 
\end{itemize}

In the gauge-fixed version  the action is quite simple,
and versions for different p's can be obtained from a single
10D Action by dimensional reduction:
\begin{equation}
 - \int d^{10} \sigma \sqrt{- {\rm det}\, \left(
\eta_{\mu\nu} + F_{\mu\nu} - 2 \bar\lambda \Gamma_{\mu}
\partial_\nu \lambda + \bar \lambda \Gamma^\rho \partial_\mu \lambda
\cdot \bar\lambda \Gamma_\rho \partial_\nu \lambda \right)}.
\end{equation}

Here $\lambda$ is the one surviving Majorana-Weyl spinor, with
the second one having been set to zero. The number of physical degrees
of freedom is  $8$ for the gauge field, and for the fermions
the 32 original components are reduced in half by local kappa 
symmetry and again in half by the equations of motion.

Among the symmetries of the gauge-invariant action are the
global translations \mbox{$\delta  X^m = a^m $.} Decomposing these
into translations $a^{\mu}$ tangent to the brane and $a^i$
normal to the brane, we have
\be
\begin{array}{rll}
\delta  X^{\mu}={}&a^{\mu} + \xi^{\rho}\partial_{\rho}X^{\mu}&%
{}= a^{\mu}+\xi^{\mu}= 0~,\\
\delta  \phi^{i}={}&a^{i} + \xi^{\rho}\partial_{\rho}\phi^{i}&{}=
a^i-a^{\rho}\partial_{\rho}\phi^{i}~.
\end{array}
\ee
In these equations a compensating general coordinate transformation
is used, with a parameter $ \xi^{\mu}=-a^{\mu} $, in order to 
preserve the $X^{\mu}=\sigma^{\mu} $ gauge.
One can also infer the induced transformation of the gauge field
$A_{\mu}$ :
\be
\delta   A_{\mu}=-a^{\rho}F_{\rho\mu}= -a^{\rho}\partial_{\rho}A_{\mu}+
a^{\rho}\partial_{\mu}A_{\rho}
\ee
This broke down into two pieces: first, the translation proper;
second, a gauge transformation with parameter $\chi$, such that 
$\chi=a^{\rho}A_{\rho}$.

In order to clarify the meaning of these formulas it is useful to specialize
to the cases of completely normal and completely tangent translations.
For small translations $a^i$, $\xi^{\mu} = 0$ and thus:
\be
\delta  X^{\mu}=0~,~~\delta  \phi^{i}=a^i~,~{\rm and}~~~\delta   A_{\mu}=0~~.
\ee
This is just a trivial shift, but nonetheless a 
significant symmetry of the theory:
it corresponds to spontaneously broken
(by the presence of the brane) translational symmetry normal
to the brane, and $\phi^i$ are the associated Goldstone bosons.

The second case, in this context, is the unbroken tangent
translations $a^{\mu}$. The formulas read:
\be
\delta  X^{\mu}=0~,~~{\rm thus}~~ \xi^{\mu} = -a^{\mu}~.
\ee
The dynamical fields transform as
\be
\nonumber 
\begin{array}{l}
\delta  \phi^{i}~=-a^{\mu}\partial_{\mu}\phi^{i}~,\\
\delta   A_{\nu}=-a^{\mu}\partial_{\mu} A_{\nu}+a^{\mu}\partial_{\nu} A_{\mu}~.
\end{array}
\ee
We will use an ansatz inspired by these formulas to describe tangent
fluctuations of the F-string/D-brane configuration in section \ref{neumann}.
In that case the translation parameter $a^{\mu}$ is no longer constant,
but is position dependent. In fact, it should be clear from the form of 
the previous expressions that $a^{\mu}$ is presumed small. 
Nevertheless the idea of a compensating
general coordinate transformation remains useful to make sure 
that the fields, i.e. $\phi$ and $A_{\mu}$ represent the desired 
bending of the surface of the brane.

For completeness we also quote the SUSY transformations of the 
supersymmetric version of B-I theory
\begin{eqnarray}
\Delta \bar\lambda &=& \bar\epsilon_1 + \bar\epsilon_2 \zeta^{(p)} + \xi^\mu
\partial_\mu \bar\lambda \nonumber \\
\Delta \phi^i &=& (\bar\epsilon_1 - \bar \epsilon_2 \zeta^{(p)}) \Gamma^i
\lambda + \xi^\mu \partial_\mu \phi^i \nonumber \\
\Delta A_\mu &=& (\bar\epsilon_2 \zeta^{(p)} - \bar\epsilon_1) (\Gamma_\mu +
\Gamma_i \partial_\mu \phi^i)\lambda \nonumber \\
& &  + ({1\over 3} \bar\epsilon_1 - \bar\epsilon_2 \zeta^{(p)}) \Gamma_m \lambda
\bar\lambda \Gamma^m \partial_\mu \lambda + \xi^\rho \partial_\rho A_\mu -
\partial_\mu \xi^\rho A_\rho~, \label{ptrans}
\end{eqnarray}
where $\xi$, as explained above, is the parameter of a compensating
general coordinate transformation that restores the static gauge,
i.e. $\delta \bar\theta_2 = 0$ and $\delta X^\mu = 0$ under 
these transformations. From these requirements it follows that
\begin{equation}
\xi^\mu = (\bar\epsilon_2 \zeta^{(p)} - \bar\epsilon_1) \Gamma^\mu \lambda.
\label{xivalue}
\end{equation}
Here and above $\zeta^{(p)}$ represents a matrix, exact form of which can be 
found in \cite{schwarz}, suffice it to say that key property of it is
$\zeta^{(p)}\tilde{\zeta}^{(p)}=\tilde{\zeta}^{(p)}\zeta^{(p)}=1$.

\section{Non-Abelian Generalizations of B-I }
\label{nonabelian}
The efforts to find a generalisation of B-I action to the
non-Abelian case are motivated by the need to effectively
describe N coincident branes, where the $U(1)^N$ symmtery
is enhanced to $U(N)=SU(N) \times U(1)$. This can be understood
as the effect of the massive vector bosons (open strings connecting
the different branes) becoming light/massless as the branes are moved
closer.
 
The earliest proposals entail formally expanding the ordinary B-I
Lagrangian in powers of $F$, then taking a simple \cite{hagiwara}, 
an anti-symmetrized  \cite{argyres}, 
or a symmetrized \cite{tseytlin} trace over the group indices. This is necessary
because if the Determinant is understood to be taken over the Lorentz indices,
then the only way to deal with the color index is to expand the expression
under the square root, and only then deal with the color index. Below
we present a recent proposal \cite{park} to compute the determinant
with respect to combined Lorentz and color indices.

The matters are further complicated by the difficulty to disentangle
the derivative terms~$\partial F$ which arise in the perturbative expansion
from the commutator terms, as can be seen from the following identity
for a commutator of covariant derivatives:
\be
[ D_m, D_n ] F_{kl} = [ F_{mn} , F_{kl} ] ~.
\ee
Tseytlin proposed to resolve this by treating all terms that can be 
written as a commutator to be part of the ``derivative dependent'' 
part of the effective action. In this way the $F^3$ terms in
the open bosonic string effective action can be dropped, and the $F^4$
terms differ with the superstring case by only a commutator term also.
In fact, these terms can be written as
\be
{\rm Tr} \left( {4 \over 3} i \ap F_{mn}[F_{ml},F_{nl}] \right) + 
2 \ap^2 {\rm Tr} \left(F^{mn}F^{rl} [F_{mn}, F_{rl} ] \right)
\ee
and thus belong according to his definition to the covariant derivative 
part of $L_{eff}$. Tseytlin then proceeds to give a general argument 
for taking the symmetrized trace in the formal expansion of the 
Abelian B-I action. He gets, to quadratic order
\bea
\nonumber
L_{NBI}={\rm STr} \sqrt{ det( \delta_{mn} +F_{mn})} =
{\rm STr} \left[ F^2_{mn} + 
{1 \over 2} (2 \pi \ap)^2 ( F^4- { 1 \over 4} (F^2)^2) + O(\ap^3) \right]\\
\nonumber
={\rm Tr} \Big[ F_{mn}^2 - { 1 \over 3} (2 \pi \ap)^2 ( F_{mn} F_{rn} F_{ml} F_{rl}
+ {1 \over 2}  F_{mn} F_{rn} F_{rl} F_{ml} -\\
\nonumber
{ 1 \over 4} F_{mn}F_{mn}  F_{rl}  F_{rl} - 
{ 1 \over 8} F_{mn}F_{rl}  F_{mn}  F_{rl} + O(\ap^4) \Big]
\eea

In the paper by W. Taylor IV and A. Hashimoto \cite{wati} this 
functional was checked against e.g. the excitation spectrum of open
strings on branes intersecting at an angle. They show that this action can
be used to resolve some but not all of the discrepancies.

A more recent proposal by J. Park \cite{park} is in some sense very
natural, similar ideas were circulating in early `97 , i.e. to
combine the spacetime and the group indices into one and then compute a 
single determinant. This is in fact typical of YM effective actions:
see for example the original computation of the pure YM effective action by
G. Savvidy \cite{george}. There too it is computed by exponentiating
the matrix in order to trade the determinant for a trace (see below).
F and G become matrices with a double index as follows:
\begin{equation}
F_{\mu\alpha,\,\nu\beta}\equiv F^{a}_{\mu\nu}T^{a}_{\alpha\beta}
\end{equation}
and
\begin{equation}
g_{\mu\nu}\longrightarrow G_{\mu\alpha,\,\nu\beta}\equiv 
g_{\mu\nu}\delta_{\alpha\beta}~.
\end{equation}

After defining a convenient subsidiary matrix ${\cal F}\equiv G^{-1}F$,
Park proposes the following  non-Abelian generalization of the Born-Infeld    
Lagrangian
\begin{equation}
\begin{array}{ll}
L&=2N\kappa^{-2}\left(
\left(\Det(G+\kappa F)\right)^{\frac{1}{2N}}-\sqrt{|g|}\right)\\
{}&{}\\
{}&=2N\kappa^{-2}\sqrt{|g|}\left(\mbox{exp}
\left(\sum^{\infty}_{n=1}\,\frac{(-1)^{n-1}}{n}
\textstyle{\frac{1}{2N}}\kappa^{n} {\rm Tr} {\cal F}^{n}\right)-1\right)\\
{}&{}\\
{}&=\sqrt{|g|}\left(-\frac{1}{2} {\rm Tr} {\cal F}^{2}+\frac{1}{3}\kappa {\rm Tr} {\cal F}^{3}
-\frac{1}{4}\kappa^{2}{\rm Tr} {\cal F}^{4}+\frac{1}{16N}\kappa^{2}({\rm Tr} {\cal F})^{2}
+\cdots\right)
\end{array}
\label{NBI}
\end{equation}
where $\kappa=2\pi\ap$ is a coupling constant.
The $2N$-th order root is inserted in order to insure that
for large $F$ the action is bounded from above by a fixed power
of $F$, equal to $p+1$ and independent of $N$. The second, related
reason is that $\Det \vert G \vert = \vert g \vert^N $.

In the $SU(2)$ case it is possible to compute 
the determinant in closed form. In Euclidean space:

\begin{equation}
\begin{array}{ll}
\L_{\rm {E}}&=4\kappa^{-2}\left[1-\left(1
-\textstyle{\frac{1}{4}}\kappa^{2}{\rm Tr} F^{2}
+\textstyle{\frac{1}{6}}\kappa^{3}{\rm Tr} F^{3}
-\textstyle{\frac{1}{8}}\kappa^{4}{\rm Tr} F^{4}
+\textstyle{\frac{1}{32}}\kappa^{4}({\rm Tr} F^{2})^{2}
\right)^{\frac{1}{2}}\right]\\
{}&{}\\
{}&=4\kappa^{-2}\left[1-\left(
1-\textstyle{\frac{1}{8}}\kappa^{2}F^{a}_{\mu\nu}F^{a}_{\mu\nu}
-\textstyle{\frac{1}{24}}\kappa^{3}
\epsilon^{abc}F^{a}_{\mu\nu}F^{b}_{\nu\lambda}F^{c}_{\lambda\mu}+
\textstyle{\frac{1}{128}}\kappa^{4}
(F^{a}_{\mu\nu}F^{a}_{\mu\nu})^{2}\right.\right.\\
{}&{}\\
{}&~~~~~~\,\,\,\,\,\,\,\,\,\,\,\,\left.\left.-\textstyle{\frac{1}{32}}
\kappa^{4}
F^{a}_{\mu\nu}F^{a}_{\nu\lambda}F^{b}_{\lambda\rho}F^{b}_{\rho\mu}+
\textstyle{\frac{1}{64}}\kappa^{4}
F^{a}_{\mu\nu}F^{b}_{\nu\lambda}F^{a}_{\lambda\rho}F^{b}_{\rho\mu}
\right)^{\frac{1}{2}}\right]
\end{array}
\label{su2action}
\end{equation} 

Further in the same paper an instanton solution is
constructed through a combination of the usual ansatz with
some numerical integration. The total action of this solution
seems to have the correct value $8 \pi^{2}$.


In conclusion we would like to mention that despite certain progress, the issues
surrounding the nonabelian generalisation of the B-I action are still unclear,
and in any case actual applications to brane physics are lacking.
However we will be able to sidestep the ambiguities pointed out above
when we consider in Chapter 3 a fivebrane in the essentially non-Abelian
background of $N$ D3-branes. There the dynamics of the fivebrane is captured
by the usual abelian worldvolume gauge field, which it might be possible
to interpret as the $a-la$ t'Hooft Abelian projection of the $SU(N)$ fields.

\chapter{Applications of B-I Action to the 3-brane}
\label{3brane}

\section{Electrostatics on the threebrane}
\label{statics}
The starting point for this thesis is the paper by Callan and
Maldacena \cite{cm}, hereafter referred to as CM.
They, noting that most treatments of D-branes were done using
the linearized theory, undertook to fully account  for the
specific non-linearities of the Born-Infeld
{\it super-electrodynamics}. The static features of 
intersecting branes are determined by SUSY and BPS arguments,
however dynamical issues have to be explored using the full 
Born-Infeld action. They construct, first and foremost, the 
 solution for a fundamental string ending on a D-brane, in which
the end-point of a string manifests itself as a point Coulomb charge
in the world-volume of the brane. They also find a non-BPS solution
correspoding to a brane and an anti-brane joined by e.g. a
fundamental string.

We repeat their arguments for the construction. Starting from
a 10-dimensional B-I action and dimensionally reducing to 3+1
dimensions we need to compute the action
\be
S = - \frac{1}{g_p}\int d^{4}x
\sqrt{-\Det (G_{\mu\nu} + {F}_{\mu\nu})}~~~,
{\rm with}~~g_p=(2\pi)^3g_s
\ee
where only one of the 6 available scalars is excited. The actual
determinant that needs to be computed is
\bea
\nonumber
-{\rm Det}=\left[
\mbox{$
\begin{array}{ccccc}
-1  & E_1 & E_2 & E_3 & \PD_0\phi      \\
-E_1& 1   & B_3   & -B_2   &\PD_1\phi\\
-E_2& -B_3   & 1   & B_1   &\PD_2\phi\\
-E_3& B_2   & -B_1    & 1   &\PD_3\phi\\
-\PD_0\phi   & -\PD_1\phi & -\PD_2\phi & -\PD_3\phi &1
\end{array}$}
\right]=\\
=1+ \vec{B}^2 - \vec{E}^2 -\left(\vec{E} \cdot \vec{B} \right)^2 -
   (\partial_0 \phi)^2(1+\vec{B}^2) +(\vec{\partial}\phi)^2
\label{cmatrix}\\
   +\left(\vec{B}\cdot \vec{\partial} \phi\right)^2 - 
    \left(\vec{E}\times \vec{\partial}\phi\right)^2+
   2\partial_0 \phi\left(\vec{B}\cdot[\vec{\partial} \phi\times\vec{E}]\right)
\nonumber
\eea
This expression was computed by myself  in June 1997,
and a slightly less general version can  be found 
in the review by Gibbons \cite{gibbons}.
In CM supersymmetry analysis is used to find the required solution.
This is possible because for supersymmetric BPS configurations the value
of the B-I action coincides with the linear theory. Thus one can 
find the required solution by constructing one for the linear theory
and then checking that it satisfy the equations of motion of the
full non-linear theory.

This is done in more detail in section \ref{brane_death} for
the more general, non-BPS static configuration that corresponds to a 
threebrane and anti-brane joined by a fundamental string. The limit 
of a single brane with an infinite string attached is simply
the limit of infinite separation between the branes, or 
$A \rightarrow B$ whereby $A$ and $B$ are the integration
constants for the electric and scalar fields respectively.

In order to use BPS arguments to find solutions of the three-brane 
worldvolume gauge theory which have the interpretation of string 
ending on a three-brane, we, after CM, consider the supersymmetry 
variation of the gaugino
\be
\label{gauginovar}
{\delta \chi = \Gamma^{mn} F_{mn} \epsilon}
\ee
where $mn$ are ten-dimensional indices. This is the dimensional 
reduction decription, so that the coordinate and electromagnetic
massless excitations are both part of the same ten-dimensional 
electromagnetic field strength $F_{mn}$ as is the case in the above 
matrix (\ref{cmatrix}).

We would like to find a solution to (\ref{gauginovar}) for which
$\delta\chi=0$ for some $\epsilon$. This is not possible for a point 
charge Coulomb gauge field alone. However we can also excite one of the 
transverse scalar fields such that $\phi=A_0={c \over r}$. This
is clearly a solution of the linearized (Laplace) equation.
Taking into account that $F_{0r}=F_{09}$, the supersymmetry variation 
(\ref{gauginovar}) reduces to
\be
\label{preservedsusy}
( \Gamma^{0r} + \Gamma^{9r}) \epsilon = 0~~\Rightarrow~~
(\Gamma^{0} + \Gamma^{9}) \epsilon = 0~~\Rightarrow~~
(1+\Gamma^{09})\epsilon = 0~.
\ee
Therefore a SUSY spinor, polarized along the directions of
the worldsheet $(09)$ preserves half the supersymmetry. This is the 
same spinor as in Dabholkar and Harvey, even though here the 
situation is complicated by the simultaneous presence of the 
threebrane and the fact that the string is half-infinite. In section 
\ref{secSUSY} we present a more complete explanation of the conditions imposed
on SUSY variations in the presence of various branes. In the context
of type IIB SUGRA there is a further condition due to the presence of 
the threebrane, which can be written as
\be
\Gamma^{0123} \epsilon = i \epsilon ~.
\ee
One can find the charge quantization condition in CM, and determine
the value of $c$ in the solution as $ c=\pi g$. Thus the solution 
finally is
\be
\phi = \frac{c}{r}~,~~~~A_0=\frac{c}{r}~,~~~~
\vec{E} = -\frac{c}{r^2}\vec{e_r}~.
\label{spike}
\ee
Further, CM show that this infinite spike should be interpreted as an
infinite fundamental string. First, one can derive the Hamiltonian for 
the static configurations that we have been so far considering
\be
H = \frac{1}{g_p} \sqrt{ (1+(g_p \vec{\Pi})^{2}) (1+ (\vec{\nabla} X)^{2})-
(g_p \vec{\Pi} \times \vec{\nabla}X)^2 }~~.
\label{hamil}
\ee
Using this expression, one can confirm that the energy per unit length
of the ``string-like'' object is indeed equal to that of a fundamental
string $dE={1\over2\pi}d\phi$ or $T={1\over2\pi}$.

In the next two sections it is shown that this picture survives a more 
detailed scrutiny, by considering the dynamical behavior of this
configuration under small perturbations. Polchinski's open string boundary 
conditions (Dirichlet or Neumann) arise from the full Born-Infeld 
theory dynamically.

\section{Dynamics: Dirichlet Boundary Conditions}
\label{dirichlet}
Having presented the static B-I charge configuration, we go on to show
that small excitations which propagate along the string, in the low-energy limit
completely reflect with 
\begin{itemize}
\item Dirichlet boundary conditions for excitations completely orthogonal
to both the brane and the string, (this section)\footnote{
This section is based on CM. Similar conclusions were reached in \cite{larus}
by considering the covariant worldsheet action in the supergravity
background of the threebrane.}
\item Neumann boundary conditions for normal excitations of the string 
that are polarized along one of the directions of the brane. 
(next section \ref{neumann})
\end{itemize} 
In order to consider the first case we need to add another scalar field
into consideration. The convenient form of B-I is the dimensional reduction from
10D. The relevant determinant is
\be
\left(
\begin{array}{cccccc}
-1  & E_1 & E_2 & E_3 & 0       &\PD_0\zeta\\
-E_1& 1   & 0   & 0   &\PD_1\phi&\PD_1\zeta \\
-E_2& 0   & 1   & 0   &\PD_2\phi&\PD_2\zeta \\
-E_3& 0   & 0   & 1   &\PD_3\phi&\PD_3\zeta \\
0   & -\PD_1\phi & -\PD_2\phi & -\PD_3\phi &1&0\\
-\PD_0\zeta &-\PD_1\zeta&-\PD_2\zeta&-\PD_3\zeta &0&1
\end{array} 
\right)
\ee
Computing this and substituting the background values of the fields 
gives the following action for the dynamics
of the $\zeta$ field:
\be
L= - \int d^4 x \sqrt{ 1 - (\partial_0 \zeta)^2 + 
\left(1+{c^2\over r^4}\right) (\partial_i\zeta)^2 }
\ee
The quadratic part, relevant to low-energy excitations is thus
\be
L_q = \int d^4 x \left[  (\partial_0 \zeta)^2 - 
\left(1+{c^2\over r^4}\right) (\partial_i\zeta)^2 \right] 
\ee
The resulting equation of motion is
\be
-\left(1+{c^2\over r^4}\right)\partial_t^2\zeta
     +r^{-2}\partial_r\left(r^2\partial_r\zeta\right) = 0~.
\ee

In order to fully verify the correspondence with Polchinki's picture of
D-brane dynamics via open fundamental strings, we have to look into the
effective boundary condition for small fluctuations on the string (the $r\to 0$
region) imposed by the presence of the three-brane (the $r\to\infty$ region).
In other words, we want the S-matrix connecting the two asymptotic regions.

We would like to look for solutions  of definite energy (frequency). With the convenient
redefinition of radius and coupling, the equation becomes
\be
\left( 1+{\kappa^2\over x^4}\right)\zeta+ 
x^{-2}{d\over dx}\left(x^2{d\over dx}\zeta\right)=0 \qquad{\rm with}~~
x=\omega r~,~~{\rm and}~~~  \kappa = c\omega^2~.
\ee
At this point the problem gets a unified dependence on energy and coupling
through the parameter $\kappa=c\omega^2=\pi g_s \omega^2$.
The equation has the interesting symmetry $r \leftrightarrow \kappa/r$ with 
$ \zeta \leftrightarrow \zeta \cdot r $. This seems to imply that the interior
asymptotic region, i.e. the string, is completely equivalent, or dual,
to the outer region, i.e. the brane. Notice also the $ 1+ {1 \over r^4 } $ factor,
familiar from the SUGRA solution for the 3-brane. 
See also \cite{larus} for first hints of distance/radius duality in AdS. 
In order to treat the two asymptotic regions on an equal 
footing and to elucidate the fact
that excitations "slow down" as $r\to 0$ it turns out to be convenient to
introduce a new radial coordinate which measures the distance along the 
surface of the spike 
\be
\xi(x) =\int\limits_{\sqrt{\kappa}}^x dy \sqrt{1+{\kappa^2\over y^4}}
\ee
and a new wavefunction designed such that the Laplacian preserves its usual form
in this coordinate
\be
\tilde\zeta = \zeta \cdot x\left(1+{\kappa^2\over x^4}\right)^{1/4}~.
\ee
The coordinate behaves as $\xi \sim r$ in the outer region 
($r\rightarrow\infty$)
and $\xi \sim -\kappa/r$ on the string ($r\rightarrow 0$). 
The exact symmetry of the equation
$r \leftrightarrow \kappa/r$ goes over to $ \xi \leftrightarrow - \xi$. 
The equation, when written in this coordinate becomes just the free 
wave equation, plus a narrow symmetric potential at $\xi \sim 0$
\be
-{d^2\over d\xi^2}\tilde{\zeta}+\frac{5\kappa^2}
{(x^2+\kappa^2/x^2)^3}\tilde{\zeta}=\tilde{\zeta}~.
\ee
The asymptotic wave functions can be constructed as plane waves in $\xi$, 
$$
\tilde{\zeta}=e^{\pm i \xi(x)} \qquad 
\zeta(x) = (x^4 +{\kappa^2})^{-1/4} e^{\pm i \xi(x)}~,
$$
or in the various limits:
\bea
\nonumber
x\rightarrow ~0~~~~\zeta \sim ~~e^{\pm i \xi(x)}~,\\
\nonumber
x\rightarrow \infty~~~~\zeta \sim {1 \over x}e^{\pm i \xi(x)}~.
\eea
These formulae give us the asymptotic wave function in the regions 
$ \xi \rightarrow \pm \infty$, 
while around $\xi = 0~(x=\sqrt{\kappa})$ there is a symmetric repulsive potential
which drops very fast $ \sim 1/\xi^6$ on either side of the origin.
The scattering is described by a single dimensionless parameter 
$\kappa=c \omega^2$, and in the limit of small frequency $\omega$ and/or coupling 
$c=\pi g_s$ the potential becomes narrow and high, and  can be 
replaced by a $\delta$-function with an equivalent  area 
under the curve:
\bea
\nonumber
{\cal U}=\int V(\xi) d\xi=
\int\limits_0^{\infty}\frac{5\kappa^2}{(x^2+\kappa^2/x^2)^3}~\sqrt{1+{\kappa^2\over x^4}}~ dx=\\
\frac{5\Gamma(5/4)^2}{3\sqrt{\pi\kappa}}={.77 \over \sqrt{\kappa} }
\eea
Thus,
\be
\label{potlim} 
\begin{array}{lll}
V(\xi) \sim &{.77\over\sqrt{\kappa}} \delta(\xi) \qquad &\kappa\to 0\\
V(\xi) \sim  &0  \qquad &\kappa\to\infty 
\end{array}
\ee
Therefore the scattering matrix becomes almost
independent of the exact form of the potential. We are interested in the behavior
of the one-dimensional reflection and transmission amplitudes $R$ and $T$.
The result is that
most of the amplitude is reflected back with a phase shift close to $\pi$, thus
dynamically realizing the Dirichlet boundary condition in the low energy
limit:
\be
\begin{array}{ll}
\kappa\to 0   \qquad      &\qquad \kappa\to\infty    \\
{}&{}\\
R\to -1-2i\sqrt{\kappa}/.77 \qquad     & \qquad R\to 0     \\
{}&{}\\
T\to -2i\sqrt{\kappa}/.77 \qquad & \qquad |T|\to1
\end{array}
\label{smatrix}
\ee
We also learn from this that there is a smooth crossover from
perfectly reflecting to perfectly transparent behavior as a 
function of the energy. Moreover WKB arguments show  that
the phase shift of the transmitted wave does not tend to zero,
but to a constant of order unity.

However a more detailed analysis shows that $\delta$-function approximation is not
exactly correct, even in the limit of small $\kappa$. A better approximation
is that of a square potential, which can be adjusted with two parameters
to match both the integral of the potential and the integral of the square
root of the potential, a quantity familiar from WKB underbarrier tunneling.
The integral of the square root of the potential is
\bea
\nonumber 
{\cal S}=\int \sqrt{V(\xi)}~d\xi=
\int\limits_0^{\infty}
\sqrt{ \frac{5\kappa^2}{(x^2+\kappa^2/x^2)^3}}~~\sqrt{1+{\kappa^2\over x^4}}~ dx=\\
{\sqrt{5} \over 4} \pi=1.75...
\eea
The detailed  computation for a square potential gives the following
approximate amplitudes, in terms of above introduced parameters ${\cal S}$ and ${\cal U}$
\bea
\label{ampsquare}
\begin{array}{ll}
R\to -1&{}-{2 i {\cal S} \over {\cal U} \sinh{{\cal S} } } \\ 
{}&{}\\
T \to &{}-{2 i {\cal S} \over {\cal U} \sinh{{\cal S} } }
\end{array}
\eea
Note that this coincides with (\ref{smatrix}) for ${\cal S} \to 0$. It also 
incorporates the correct exponential falloff in the wavefunction. As calculated
above, our specific potential has ${\cal S} \sim 1.75 $. Thus the scaling
of the amplitudes with $\kappa$ is not essentially modified in this more
detailed calculation, but only the exact coefficient is affected:
\bea
\nonumber
R \to -1-i 0.92 \sqrt{\kappa}\\
T \to ~~-i 0.92 \sqrt{\kappa}  
\eea
With this we conclude the analysis of completely transverse excitations 
of the string/brane system, and proceed in the next section to
the considerably more complicated longitudinal excitations. As
we will argue in that case, there are reasons to believe that 
the coefficient computed above should in fact be exactly equal to one. 

\section{Dynamics: Neumann Boundary Conditions}
\label{neumann}
\noindent
We would like to show that certain excitations of the F-string/D3-brane 
system  obey Neumann (free) boundary conditions by considering 
the Born-Infeld dynamics of the F-string (viewed as a 3-brane cylindrically
wrapped on an $S_2$). In the paper by Callan and Maldacena it was shown that
excitations which are normal to both the string and the 3-brane behave as if
they had Dirichlet boundary conditions at the point of attachment. Here 
we show that excitations which are coming down the string with a 
polarization along a direction parallel to the brane are almost completely
reflected just as in the case of all-normal excitations, but the end of
the string moves freely on the 3-brane, thus realizing Polchinski's
open string    Neumann
boundary condition dynamically. In the low energy limit 
$ \omega \rightarrow 0$, i.e. for wavelengths much larger than the string
scale only a small fraction $ \sim \omega^4$ of the energy escapes in
the form of dipole radiation.
The physical interpretation is that a string attached to the 3-brane 
manifests itself as an electric charge, and waves on the string cause
the end point of the string to freely oscillate and therefore produce 
e.m. dipole radiation in the asymptotic outer region.

Dai, Leigh and Polchinski \cite{pol} introduced D-branes 
as objects on which strings can end, 
and required that the string have Dirichlet (fixed) boundary conditions
for coordinates normal to the brane, and Neumann (free) boundary 
conditions for coordinate directions parallel to the brane \cite{rr,pol,leigh}. 
It was
shown in \cite{cm} (see previous section \ref{dirichlet}) 
that small fluctuations which are normal to both the 
string and the brane are mostly reflected back with a 
$phase~ shift  \rightarrow \pi$
which indeed corresponds to Dirichlet boundary condition.
See also \cite{larus} and \cite{rey} for 
a  treatment of this problem, that relies on 
Nambu-Goto (or covariant) action for the   
worldsheet of the string placed in the supergravity
background of the threebrane.

In this section we will show that P-wave excitations which are coming down 
the string with a
polarization along a direction parallel to the brane are almost completely
reflected just as in the case of all-normal excitations, but the end of
the string moves freely on the 3-brane, thus realizing Polchinski's
open string    Neumann boundary condition dynamically. As we will see  
a superposition of excitations of the 
scalar $x_9$ and of the e.m. field reproduces the required behaviour,
e.g. reflection of the geometrical fluctuation with a 
$phase~ shift  \rightarrow 0$
(Neumann boundary condition)\footnote{ 
This problem was also considered in \cite{rey} where the e.m. field is 
integrated out to produce an effective lagrangian for the scalar field
only. The other essential difference with us is that 
we consider $P$-wave modes of the scalar field which describe physical 
transverse fluctuations of the F-string and not the $S$-wave modes  
which do not correspond to physical excitations of the string. }.

In addition we observe e.m. dipole radiation which escapes to infinity
from the place where the string is attached to the 3-brane. 
We shall see that in the low energy limit
$ \omega \rightarrow 0$, i.e. for wavelengths much larger than the string
scale a small fraction $ \sim \omega^4$ of the energy escapes to infinity in
the form of e.m. dipole radiation.
The physical interpretation is that a string attached to the 3-brane
manifests itself as an electric charge, and waves on the string cause
the end point of the string to freely oscillate and therefore produce
e.m. dipole radiation in the asymptotic outer region of the 3-brane.
Thus not only in the static case, but also in a more general dynamical 
situation the above interpretation remains valid.
This result provides additional support to the idea that 
the electron may be understood as the end of a fundamental string 
ending on a D-brane.

\subsection{The Lagrangian and the equations}

Let us write out the full Lagrangian (see section which contains both 
electric and magnetic fields, plus the scalar $x_9 \equiv \phi$
\bea
\nonumber
L=-\int d^4x \sqrt{\Det}~,~~~~~~~~~~~~~~~~~~~~~~~~~~~~~
~~~~~~~~~~~~~~~~~~~~~~~~~~~~~~~~~~\\
\nonumber
{\rm where}~~~\Det=1+ \vec{B}^2 - \vec{E}^2 -\left(\vec{E} \cdot \vec{B} \right)^2 -
   (\partial_0 \phi)^2(1+\vec{B}^2) +(\vec{\partial}\phi)^2\\
   +\left(\vec{B}\cdot \vec{\partial} \phi\right)^2 - \left(\vec{E}\times \vec{\partial}\phi\right)^2+
   2\partial_0 \phi\left(\vec{B}\cdot[\vec{\partial} \phi\times\vec{E}]\right) 
\label{lagrfull}
\eea
We will proceed by adding a fluctuation to the background values 
(\ref{spike}) : 
$$ \vec{E}=\vec{E}_0 + \delta \vec{E},~~\vec{B}= 
\delta \vec{B},~~\phi=\phi_0 +\eta~.$$
Then keeping only terms in the $Det$  which are linear and quadratic
in the fluctuation we will get 
\bea
\delta Det =  \delta\vec{B}^2 - \delta\vec{E}^2 - 
(\vec{E_{0}} \delta\vec{B})^2  - (\partial_0 \eta)^2 +   
(\vec{\partial}\eta)^2 \\ \nonumber
+ \left(\delta\vec{B} \vec{\partial} \phi\right)^2
- \left(\vec{E_0} \times \vec{\partial}\eta\right)^2 - 
\left(\delta\vec{E} \times \vec{\partial} \phi\right)^2 
-2\left(\vec{E_0} \times \vec{\partial}\eta\right)
\left(\delta\vec{E} \times \vec{\partial}\phi\right)\\ \nonumber
-2\left(\vec{E_0} \delta\vec{E}\right) + 2 \left(\vec{\partial}\phi \vec{\partial}\eta\right)
\eea
Note that one should keep the last two linear terms because they produce 
additional quadratic terms after taking the square root.
These terms involve the longitudinal polarization of the 
e.m. field and cancel out at quadratic order.
The resulting quadratic Lagrangian is 
\be
2L_q = \delta\vec{E}^2\left(1+(\vec{\partial} \phi)^2\right) - \delta\vec{B}^2 + 
(\partial_0\eta)^2 -
(\vec{\partial}\eta)^2\left(1-\vec{E_0}^2\right) + \vec{E_0}^2 
\left( \vec{\partial}\eta \cdot \delta\vec{E}\right)~~.
\label{full}
\ee
Let us introduce the gauge potential for the fluctuation part of the e.m.
field as $(A_0,\vec{A})$ and substitute the values of the background 
fields from (\ref{spike})
\be
2L_q = \left(\partial_0 \vec{A}- \vec{\partial}A_0\right)^2 \left(1+{1\over r^4}\right)-
       \left(\vec{\nabla}\times\vec{A}\right)^2 + (\partial_0\eta)^2 -
       (\vec{\partial}\eta)^2\left(1-{1\over r^4}\right) + 
       {1\over r^4} \left(\partial_0 \vec{A}- \vec{\partial}A_0\right)
       \cdot \vec{\partial}\eta ~~.
\label{fluctlag}
\ee
The equations that follow from this lagrangian contain dynamical 
equations for the vector potential and for the scalar field, 
and a separate equation which represents a constraint. These
equations in the Lorenz gauge
$\vec{\partial}~\vec{A}=\partial_0 A_0$ are
\bea
\label{alp}
~~-\partial_0^2 \vec{A}\left(1+{1\over r^4}\right) + \Delta\vec{A}~+~~~
{1\over r^4}\vec{\partial}\partial_0(A_0+\eta) = 0~~~~~&{} \\
\label{bet}
~~-\partial_0^2 A_0 + \Delta A_0 + 
\vec{\partial} {1\over r^4} \vec{\partial}(A_0+\eta) - 
\vec{\partial} {1\over r^4} \partial_0\vec{A} = 0~~~~~&{} \\ 
\label{gam}
~~-\partial_0^2 \eta~ + \Delta \eta~~ - 
\vec{\partial} {1\over r^4} \vec{\partial}(A_0+\eta) + 
\vec{\partial} {1\over r^4} \partial_0\vec{A} = 0~~~~~&{}
\eea
Equation (\ref{bet}) is a constraint: the time derivative of
the {\it lhs} is zero, as can be shown using the equation of motion 
(\ref{alp}).

Let us choose $A_0=-\eta$. This condition can be viewed as (an attempt to)
preserve the BPS relation which holds for the background: 
$\vec{E}=\vec{\partial } \phi$. Another point of view is that this fixes
the general coordinate invariance which is inherent in the Born-Infeld 
lagrangian in such a way as to make the given perturbation to be
normal to the surface. Of course transversality is insured automatically
but this choice makes it explicit. The general treatment of this 
subject can be found in \cite{schwarz}. See also section \ref{gen_cov}
for a quick review of the issues involved.

With this condition the equations (\ref{bet}) and (\ref{gam}) become the same, and
the first equation is also simplified:
\bea
-\partial_0^2 \vec{A}\left(1+{1\over r^4}\right) + ~\Delta\vec{A} = 0~, \\
-\partial_0^2 \eta + \Delta \eta +
\vec{\partial} {1\over r^4} \partial_0\vec{A} = 0~.
\label{seqns}
\eea
This should be understood to imply that once we obtain a solution,
$A_0$ is determined from $\eta$, but in addition we are now obliged to 
respect the gauge condition which goes over to 
$\vec{\partial}\vec{A}=-\partial_0 \eta$.

\subsection{Neumann boundary conditions and dipole radiation}
\begin{figure}
\centerline{\hbox{\psfig{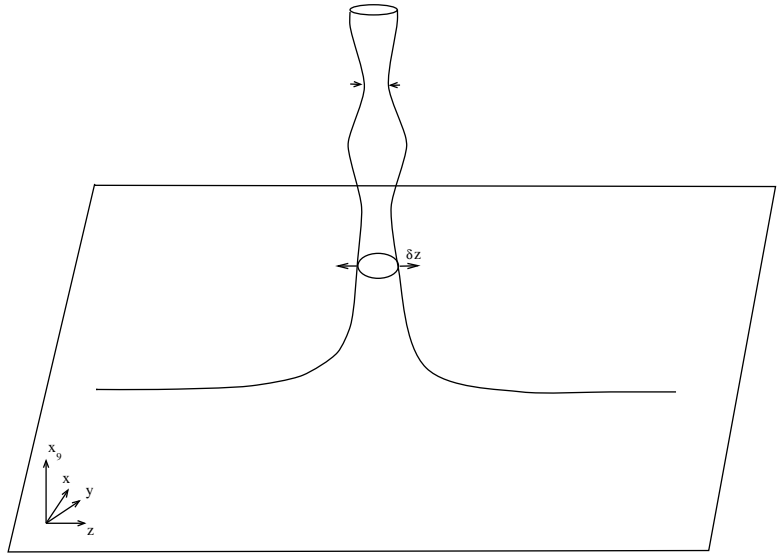}}}
\caption[braneeps]{ In order that all points on the $S_2$ section of the
tube ( which is schematically shown here as a circle) move all in
the same direction $\hat z$ by an equal distance $\delta z$, the field 
$\eta$ has to take on different values at, say, opposite points of the 
$S_2$. In effect, 
$ \eta=\delta{-1\over r}={1 \over r^2}{z \over r}\delta z$. 
If, on 
the other hand, it is taken to be an $S$-wave (as in the paper \cite{rey})
that would correspond to Fig \ref{braneeps}{\bf b},
which at best is a problematic `internal' degree of freedom of the tube.
} 
\label{braneeps}
\end{figure}

We will seek a solution with definite energy 
(frequency $w$) in the following form: $\vec{A}$ should have
only one component $A_z$, and $\eta$ be an $l=1$ spherical $P$-wave
$$
A_z= \zeta(r)~e^{-i\omega t}~~,~~~\eta={z\over r}~\psi(r) ~e^{-i\omega t}
$$
The geometrical meaning of such a choice for $\eta$ is explained in Fig \ref{braneeps},
and the particular choice of $z$ dependence corresponds to the polarization
of the oscillations along the $z$ direction of the brane.
With this ansatz the equations become
\bea
\label{oldfriend}
\left(1+{1\over r^4}\right)~\omega^2 \zeta + 
{1\over r^2}\partial_r(r^2 \partial_r \zeta)=0~~\\
{z\over r}~ \omega^2 \psi + 
{z\over r}~ {1\over r^2}\partial_r(r^2 \partial_r \psi) +
{z\over r}~ {2\over r^2}~\psi - i \omega \partial_z\left({\zeta\over r^4}\right) = 0~,
\eea
with the gauge condition becoming $\partial_r\zeta =i \omega \psi$.
It can be seen again, that the second equation follows from the first by
differentiation. This is because the former coincides with the constraint 
in our ansatz.

Therefore  the problem is reduced to finding the solution of a 
single scalar equation, and determining the other fields through
subsidiary conditions. The equation  \ref{oldfriend}
itself surprisingly turned out to be 
the one familiar  from CM \cite{cm} for the transverse fluctuations.
We refer the reader to section \ref{dirichlet} for a detailed
discussion of asymptotic wavefunctions and scattering amplitudes.

The asymptotic wave functions can be constructed as plane waves in 
the coordinate $\xi$, as introduced in Section \ref{dirichlet}, 
$$
\zeta(r) = (1 + r^4)^{-1/4} e^{\pm i \xi(r)}~,
$$
or in the various limits:
\bea
\nonumber
r\rightarrow ~0~~~~\zeta \sim ~~e^{\pm i \xi(r)}~,\\
\nonumber
r\rightarrow \infty~~~~\zeta \sim {1 \over r}e^{\pm i \xi(r)}~.
\eea
These formulae give us the asymptotic wave function in the regions 
$ \xi \rightarrow \pm \infty$, 
while around $\xi = 0~(r=1)$ there is a symmetric repulsive potential
which drops very fast $ \sim 1/\xi^6$ on either side of the origin.
The scattering is described by a single dimensionless parameter 
$\omega \sqrt{c}$, and in the limit of small $\omega$ and/or coupling 
$c=\pi g_s$ the potential becomes narrow and high, and  can be 
replaced by a $\delta$-function with an equivalent  area 
$\sim {1\over \omega \sqrt{g_s}}$ under the curve.
Therefore the scattering matrix becomes almost
independent of the exact form of the potential. The end result is that
most of the amplitude is reflected back with a phase shift close to $\pi$, thus
dynamically realizing the Dirichlet boundary condition in the low energy
limit. 

In order to obtain $\psi$ (and $\eta$) we need to differentiate
$\zeta$ with respect to $r$:
\be
i\omega\psi= {-1\over4} {4r^3\over(1+r^4)^{5/4}}e^{\pm i \xi(r)}
\pm{i\omega\over(1+r^4)^{1/4}}
\left(1 +{1 \over r^4}\right)^{1/2} e^{\pm i \xi(r)}~~.
\ee
Again it is easy to obtain the simplified limiting form:
\bea
\nonumber
r\rightarrow 0~~~~~i \omega \psi  
\sim \left(-r^3 \pm {i\omega \over r^2}\right) e^{\pm i \xi(r)}\\
\nonumber
r\rightarrow \infty~~~~i \omega \psi 
\sim \left({-1 \over r^2} \pm {i\omega \over r}\right) e^{\pm i \xi(r)}
\eea 
This brings about several consequences for 
$\psi$. Firstly, it causes $\psi$ to grow as $\sim 1/r^2$ as 
$r\rightarrow 0$. This is the correct behaviour because when  converted to
displacement in the $z$ direction, it means constant amplitude:
\be
 \eta={z\over r}\psi=\delta{-1\over r}={z \over r}{1 \over r^2}\delta z~~%
\Rightarrow ~~\delta z \sim const~~.
\ee
Secondly, the $i$ that enters causes the superposition of the 
incoming and reflected waves to become a cosine from a sine, as is
the case for $\zeta$ waves. This corresponds to a $0$ phase shift and 
implies   Neumann boundary condition for the $\eta$ wave (Figure \ref{waveeps}).

Because of the $\omega$ factor in the gauge condition we 
need to be careful about normalizations, thus we shall choose to 
fix the amplitude of the $\delta z$ (or $\eta$) wave to be independent of $\omega$. 
Then the magnitude of the e.m. field in the inner region becomes 
independent of $\omega$ as well. 
Combined with the transmission factor, proportional to 
$\omega\sqrt{c}$, this gives the following form of the 
dipole radiation field at infinity\footnote{The unit of electric charge in our notation
is $\pi g_s=e^2$. This is because of the scaling of the fields needed
to get the U(1) action with $1/e^2$ in front of it.}
\be
A_z \sim {\omega \sqrt{\pi g_s} \over r } \zeta_0 e^{-i \omega t}=
{\omega e \zeta_0\over r } e^{-i \omega t}={\dot{d} \over r }
\label{dipolerad}
\ee
In order to make a comparison with Thomson formula 
$I={4\over 3}\omega^4 e^2 A^2$ for the total
power emitted by an oscillating dipole, we note that the agreement
is guaranteed if the exact coefficient
in (\ref{dipolerad}) is equal to one. As shown at the end of the previous
section, our approximate computation of the transmission amplitude
gives this coefficient as $\sim .92$. Thus it is plausible
that the magnitude of emitted power is in fact exactly equal to the
elementary expectation from ordinary electrodynamics.
\begin{figure}
\centerline{\hbox{\psfig{figure=\path wave.eps,height=16cm,angle=0}}}
\caption[waveeps]{ The figure \ref{waveeps}{\bf a} depicts the scattering of the 
$\zeta$ wave. Note the discontinuity in the derivative which is 
proportional to $1/\omega$. Figure \ref{waveeps}{\bf b} shows the scattering of
the $\eta$ wave. Being the derivative of $\zeta$, it 
results in a discontinuity of the function itself,
making it into a cosine, which means free (Neumann) 
boundary condition at $\xi =0$.
} 
\label{waveeps}
\end{figure}

In conclusion, we need to analyze the outgoing scalar wave. This wave has 
both real and imaginary parts, the former is from differentiating the phase,
while the latter is from the prefactor. The imaginary part is $\sim 1/r^2$
which drops faster than radiation. The real part 
does contribute to the radiation at spatial infinity,
as can be shown from the integral of the energy density
$\int (\partial_r \eta)^2 d^3r \sim 
\int \omega^4/r^2 \cdot 4\pi r^2 dr \sim \omega^4$.
This is not altogether surprising, as we are dealing with a
supersymmetric theory where the different fields are tied together.
Thus the observer at spatial infinity will see both an
electromagnetic dipole radiation field and a scalar wave. 
Interestingly, the direction dependences of the two conspire
to produce a spherically symmetric distribution
of the energy radiated.

The problem of longitudinal fluctuations was treated in \cite{rey}, though 
not in a completely satisfactory way. There the scalar field was taken to be
an $S$-wave, which as should be apparent from Fig \ref{braneeps}{\bf b}, does not correspond 
to the string oscillating as a whole. In addition the electromagnetic field
was effectively integrated out, thus one cannot obtain the dipole 
radiation at spatial infinity.

\section{Brane-anti-Brane annihilation}
\label{brane_death}

\vspace{5 mm}

\noindent

We revisit the solution of Born-Infeld theory which
corresponds to a D3-brane and anti-brane joined by a
(fundamental) string. The global instability of this configuration
makes possible the semiclassical tunneling 
into a wide, short tube which keeps expanding
out, thus annihilating the brane. This tunneling is suppressed
exponentially as $\exp\{-\frac{S_{cl}}{g}\}$,
where $S_{cl}$ would be the action of the euclidean bounce solution,
the construct that describes under-barrier tunneling
in the WKB approximation. The attraction between
the branes causes them to approach and annihilate 
at a finite distance $D_{min}$,
where the potential barrier disappears.
For large separations D, the energy of the solution at the top of the barrier, the sphaleron, goes like
$\sim D^3$ , while the energy of the string is proportional
to its length D. 

Callan and Maldacena \cite{cm} considered among other
configurations the D3-brane and anti-brane joined by a (fundamental) string
in the framework of Born-Infeld theory. The string is made of a D3-brane,
wrapped aroung $S^2$ sphere. When looked from some distance, such an object
does not appear to carry RR charge as a whole,but is rather like a RR dipole,
and has energy per unit length proper to the fundamental string. 

Such a configuration is only quasi-stable,
since globally it is possible to lose energy by making the throat very wide:
if it had radius R, the change in energy is mostly due to tension\footnote{
The second term has the same origin as in \cite{emparan}. The two
branes act as a sort of capacitor to create a uniform bulk 3-form RR field to which
the cylindrically wrapped brane couples. The only difference is dimensionality:
in case of the 2-brane the potential goes as $-R^2$, and for the 3-brane
it's $\sim - R^3$.} and goes like $\sim R^2\cdot D - R^3$ 
and is arbitrarily negative.
However, there is a potential energy barrier
and one needs to construct the bouncing euclidean solution in order to
address the problem in a complete way. In CM \cite{cm} it was attempted
to approach the problem by dropping the contributions due to the electric field, 
in that case the
lagrangian is Lorentz invariant with respect to r,t and it is possible to
construct some approximation to the bouncing solution.

 In this section we will compute exactly the energy of the string and sphaleron
solution (the unstable static solution at the top of the potential barrier).
This will allow to conclude that the tunneling rate in fact goes to infinity
when the branes approach each other but still are at a finite distance
$\sim l_s\sqrt{g_s}$.

\subsection{The Two Static Configurations}

We will review the construction of the relevant solution from CM \cite{cm}.
Similar solutions were also considered by Gibbons in \cite{gibbons}.

Consider the case when the worldbrane gauge field is purely electric
and only one transverse coordinate X is excited. The worldbrane action
reduces to
\be
L = -\frac{1}{g_p} \int d^{4}x \sqrt{ (1-\vec{E}^{2}) (1+ (\vec{\nabla} X)^{2})
+ (\vec{E} \vec{\nabla} X)^{2} - {\dot{X}}^{2}}~,
\label{langrangian}
\ee
where $g_p=(2\pi)^3g_s$, and $g_s$ is the string coupling ($\alpha^{\prime} =1$).

The canonical momentum associated with $\vec{A}$ is
\be
g_p \vec{\Pi}=  \frac{ \vec{E}(1+ (\vec{\nabla}X)^{2})-\vec{\nabla}X
(\vec{E} \vec{\nabla} X )  } { \sqrt{ (1-\vec{E}^{2})
(1+ (\vec{\nabla} X)^{2})
+ (\vec{E} \vec{\nabla}X)^{2} - {\dot{X}}^{2}} }~.
\label{momentum}
\ee
The constraint equation is $\vec{\nabla}\vec{\cdot\Pi}=0$. The Hamiltonian is

\be
H = \frac{1}{g_p} \sqrt{ (1+(g_p \vec{\Pi})^{2}) (1+ (\vec{\nabla} X)^{2})-
(g_p \vec{\Pi} \times \vec{\nabla}X)^2 }~~.
\label{ham}
\ee

We are looking for the most
general static, spherically symmetric solution. The equation for X, which
follows from varying the energy, after setting $\dot{X}=0$ is
\be
\vec{\nabla}  \frac{ (1-\vec{E}^{2}) \vec{\nabla}X +
\vec{E}(\vec{E} \vec{\nabla}X) }{ \sqrt{ (1-\vec{E}^{2})
(1+ (\vec{\nabla}X)^{2})
+ (\vec{E} \vec{\nabla }X)^{2} - {\dot{X}}^{2}} }  =0~.
\label{xeqn}
\ee

From (\ref{momentum}) it follows that $g_p \vec{\Pi}= \frac{A \hat{r}}{r^2}$,
and from (\ref{xeqn})
$ \frac{\vec{\nabla} X}{\sqrt{1- E^2 + \nabla X^2 }}= \frac{B \hat{r}}{r^2}$.
Here A and B are integration constants.

The expression (\ref{momentum}) now simplifies to 
$$ g_p \vec{\Pi} = \frac{\vec{E}}{\sqrt{1-E^2 + \nabla X^2} }~.$$

The solution for the coordinate and electric field is
\be
\vec{\nabla} X=\frac{B\hat{r}}{\sqrt{r^4 + A^2 - B^2}}~~~~~~~~
\vec{E} = \frac{A\hat{r}}{\sqrt{r^4 + A^2 - B^2}}~.
\label{solution}
\ee

One can view this solution as a way to minimally break
supersymmetry, instead of $E=\nabla X$ we have $E=\frac{A}{B}\nabla X$.
In principle, A should be quantized as electric charge in units of $\pi g$.
We will be interested in $B > A$, in this case the resulting configuration
will be the 3-brane and anti-brane joined by a smooth throat. To see
this, one needs to explicitly exploit the geometry by finding X:
\be
X(r)= B \int_{r}^{\infty} \frac{ds}{\sqrt{s^4 - r_0^4}}~. 
\label{shape}
\ee

Here $ r_0^4 = B^2 - A^2 $. We have set $X(\infty)=0$, e.g. far away
the brane is flat and is at zero coordinate in the transverse direction.
$X(r_0)$ is finite, but $X^{\prime}(r_0)$ is infinite: the throat becomes
vertical at that radius. This can be continued back out through $r_0$
to give the two branes. Branes possess orientation and continuing it through
the throat we see that the new brane is of the opposite orientation: 
an antibrane.

The relations between A,B and $r_0$, $X(r_0)=D/2$ can be solved
to express $r_0$ and $B$ in terms of D and A: 
\be
D/2= c \sqrt{\frac{A^2}{r_{0}^{2}} + r_{0}^{2}} ~~~
{\rm and}~~~~~ B^2= A^2+ r_{0}^{4}~, 
~~~~~~~{\rm where}~~~ c = \int_1^{\infty} \frac{dz}{\sqrt{z^4-1}}~.
\label{dbeqn}
\ee
In the limit of large D
the two possible radii at the throat are $r_0 \sim D/2c~,$ and $~A~2c/D $.
A remark is in place here that the minimal separation for which a real root
exists is $D_{min} = 2c\sqrt{2A}$.

Knowing the energy function  $$H = \sqrt{(1+ \nabla X^2)(1+ g\Pi^2)}$$ allows
to compute the energies of the solutions exactly,

\be
E_{tot} = \frac{1}{g_p}\int_{r_0}^{\infty}
\sqrt{1+ \frac{A^2 + r_{0}^{4}}{r^4 - r_{0}^{4}}}~~
\sqrt{1+\frac{A^2}{r^4}} ~~4\pi r^2~ d r~~.
\label{energy}
\ee

At this point the temptation to make the problem completely dimensionless
becomes irresistible. Let me introduce the parameters 
$\mu=\frac{D}{2c\sqrt{A}},
~~~r=z\sqrt{A} $. We are now measuring length in units of $\sqrt{A}$, energy in
$A^{3/2}$, and  also $r_0 = \xi \sqrt{A}$:

\be
E = \frac{1}{g_p}\int_{\xi}^{\infty} \sqrt{1+ \frac{1 +
\xi^4}{z^4 - \xi^4}}~~
\sqrt{1 + \frac{1}{z^4}} ~~4\pi z^{2}~ dz~~.
\label{dimless}
\ee
Equations (\ref{dbeqn}) can be written as $\mu^2= \xi^2 + 1/\xi^2$. 
We define the two positive roots of this equation as $\xi_1$ and $\xi_2$.
One can see that $\xi_1 \cdot \xi_2 = 1$. 
The minimum separation is $\mu_{min} = \sqrt{2}$,
at which point the two roots of the quadratic equation become degenerate:
$\xi_1 = \xi_2 = 1$.

After some manipulations with the energy integral, and a variable change \\
$y=\xi / z$, we get
\be
\frac{g_p}{8\pi}E = \xi^3 \int_{0}^{1} \frac{d y}{y^{4}\sqrt{1-y^4}} +
\frac{1}{\xi} \int_{0}^{1} \frac{d y}{\sqrt{1-y^4}}~~.
\label{answer}
\ee

 We have not dealt yet with the volume infinity which manifests
itself in the fact that the first of these integrals is divergent at 
$y \rightarrow 0$.
Regularize it by subtracting $1/y^4$ from the integrand : if one were to
go back, it is exactly equivalent to computing the energy with respect
to the configuration when the two branes are flat, parallel and not joined
by any string. Denote the integrals in (\ref{answer}) as u and v, 

\begin{eqnarray}
u= \int_{0}^{1} \frac{d y}{y^4}\left(\frac{1}{\sqrt{1-y^4}}- 1\right) -
\int_{1}^{\infty} \frac{d y}{y^4}=
-\frac{1}{3} + \sum_{n=1}^{\infty} \frac{1}{4n-3}\frac{(2n-1)!!}{n!~ 2^n}
\nonumber\\
\begin{array}{lll}
u={}&{}& \frac{\sqrt{\pi}}{4} \frac{\Gamma(-3/4)}{\Gamma(-1/4)}=.43701...
\nonumber\\
v={}&  \int\limits_{0}^{1} \frac{dy}{\sqrt{1-y^4}}={}& \sqrt{\pi}
\frac{\Gamma(5/4)}{\Gamma(3/4)}=1.31103...
\end{array}
\label{numerics}
\end{eqnarray}
The answers are not surprising, since both integrals are generalized
B functions, and the second one is in fact the quarter period of the
Jacoby elliptic functions. One can also show, by using the functional
properties of the Gamma function, that $3u=v$. 
Also, constant c from (\ref{dbeqn})
is also related, in fact $c=v$. The energies of the string
and the sphaleron become, respectively
$$
E_{string}= u ( \xi^{-3} +3\xi)~~~~~ E_{sph}= u ( \xi^3 + 3\xi^{-1} )~~.
$$
Note, that we have now taken the larger of the two roots to be $\xi$, the other
root being $1/\xi$. One can even write a relation between the energies of 
the string and sphaleron solutions  $E_{string}(1/\xi)=E_{sph}(\xi)$. 
Even though this relation is formal,
(each function is defined only for arguments larger than one) 
it might be of significance in the future. The energy of a long string
after fully substituting the dimensionful units becomes 
$$ E_{string}= \frac{4\pi}{g_p} D A = 
\frac{4\pi}{8\pi^3g}D \pi g= \frac{1}{2\pi}D~~.$$
\begin{figure}
\centerline{\hbox{\psfig{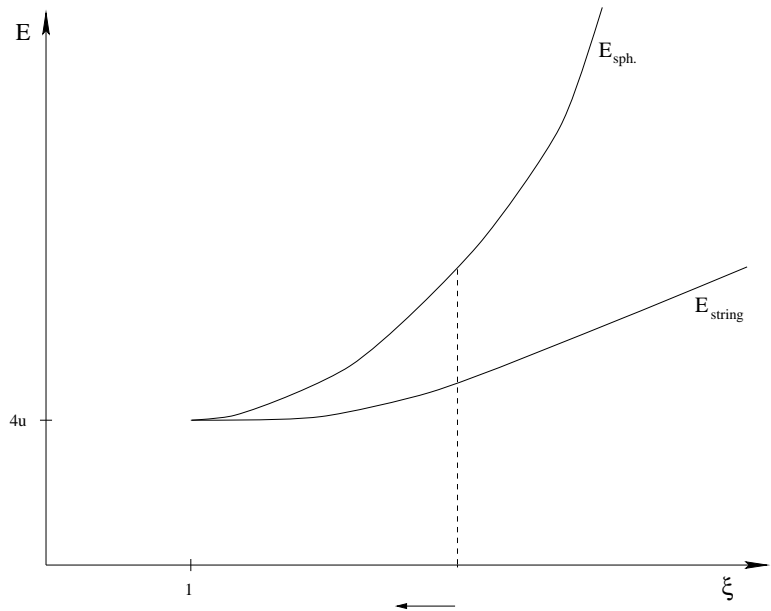}}}
\caption[fig1]{Energy vs. $\xi$. \newline {\it ~~Height of the barrier}
 = $E_{sph}-E_{string} \rightarrow 0$    } 
\label{energyeps}
\end{figure}
This reproduces the correct tension of the fundamental string.
The slope of the energy function is zero at
the minimum possible value of $\xi=1$ (Fig \ref{energyeps}).

One  further interesting elaboration  is possible. The energy surface in the
parameter space of $r_0$ and D (or $\mu, \xi$) apparently has a kink well known from
catastrophe theory. The point at which the kink first appears corresponds to
$\mu^2=2$ and $\xi=1$. For $\mu^2 < 2$  the  energy is monotonous in $\xi$, and
there is no static solutions,
otherwise it has one minimum and one maximum, corresponding to the string
and the sphaleron respectively.  

\subsection{Annihilation by Tunneling.}

Let us recall that two branes, a 3-brane and an anti-brane, are going
to gravitationally attract, even though weakly, causing them to move closer and
to eventually annihilate. The tunneling through the potential barrier,
the sphaleron being the unstable solution at the top, will be
exponentially weak \cite{cm}, but as the branes move closer $\mu$ and $\xi$
become smaller. This makes the tunneling
easier in a twofold way: by making the barrier both narrower (distance between
$\xi$ and $1/\xi$ becomes smaller), and shallower, with the energy at the top decreasing
to become equal to the energy of the string.
As $\xi \rightarrow 1$, the rate of tunneling rate increases indefinitely,
and at $\xi=1$ the metastable string state ceases to exist. This is an
interesting example of physical continuity: before the state disappeared as we
change the parameter (D or $\xi$), its physical width due to tunneling had to become
infinite. 

 Unfortunately, even though these conclusions are self consistent, 
they may still be physically
incomplete due to the fact that as branes approach to planckian distance
new nonperturbative phenomena of annihilation may kick in. One possible
way out of this difficulty is to crank up the string coupling, this leads into
unknown territory too but our quasi-classical tunneling 
may indeed be the dominant mode of annihilation in that regime.


\chapter{Born-Infeld Action and AdS/CFT Correspondence}
\label{AdS}

In this chapter we will construct solutions of the Born-Infeld action for a 
D5-brane in the background of a stack of $N$ D3-branes. The original paper 
\cite{baryon} is by C.G. Callan, A.Guijosa and myself, and I have freely
used the text of that paper here.
By building on some recent work of Imamura 
\cite{Imamura}, we can find BPS-saturated solutions which presumably correspond
to exact solutions of string theory. Using the general approach of 
\cite{cm,gibbons}, 
these solutions use D5-branes wrapped in various ways to describe branes and 
strings attached to each other. The primary object we construct this way is a 
D5-brane joined to $N$ D3-branes by a bundle of fundamental strings. Our 
solutions give a detailed description of the creation of these 
strings as the fivebrane is dragged across the threebranes.

As has been pointed out by various people \cite{wittenbaryon,groguri}, in the 
context of the anti-de~Sitter/conformal field theory ($AdS$/CFT)
correspondence there are general reasons to expect $N$ 
fundamental strings to join together on a D5-brane wrapped on a
five-sphere in the throat region of the threebrane geometry (i.e. the $AdS$ 
geometry). This is the string theory counterpart to the gauge theory 
$SU(N)$ baryon vertex, representing a bound state of $N$ external quarks.
The baryon vertex has been studied in \cite{baryonsugra}, where the 
strings and the fivebrane are 
described in terms of separate Nambu-Goto actions. 
That approach ignores the worldvolume gauge field on the fivebrane. 
Its inclusion leads to the Born-Infeld action, which 
allows a unified description of the fivebrane and the strings. When 
restricted to the $AdS$ background, our solutions provide an explicit 
string theory representation of the baryon vertex. 
The Born-Infeld action for the worldbrane dynamics of the 
fivebrane in a threebrane background is an accessible and instructive way
to go after the energetics of this problem.

In the second part of this Chapter we
study baryon configurations in large $N$ non-supersymmetric
$SU(N)$ gauge theories, applying the $AdS$/CFT correspondence.
This part is based on a paper \cite{nonextrP} by C.G. Callan, A. Guijosa,
O. Tafjord and myself.
Using the D5-brane worldvolume theory in the near-horizon geometry
of non-extremal D3-branes, we find embeddings which describe
baryonic states in three-dimensional QCD. In particular, we
construct solutions corresponding to a baryon made of $N$ quarks,
and study what happens when some fraction $\nu$ of the total
number of quarks are bodily moved to a large spatial separation
from the others. The individual clumps of quarks are represented
by Born-Infeld string tubes obtained from a D5-brane whose
spatial section has topology ${\bf R}\times {\bf S}^4$.
They are connected by a confining color flux tube, described
by a portion of the fivebrane that runs very close and parallel
to the horizon. We find that this flux tube has a tension with a
nontrivial $\nu$-dependence (not previously obtained by other 
methods).

\section{Baryons in AdS}
\label{AdS_baryon}
In this chapter we will construct solutions of the Born-Infeld action for a 
D5-brane in the background of a stack of $N$ D3-branes. 
By building on some recent work of Imamura 
\cite{Imamura}, we can find BPS-saturated solutions which presumably correspond
to exact solutions of string theory. Using the general approach of 
\cite{cm,gibbons}, 
these solutions use D5-branes wrapped in various ways to describe branes and 
strings attached to each other. The primary object we construct this way is a 
D5-brane joined to $N$ D3-branes by a bundle of fundamental strings. Our 
solutions give a detailed description of the creation of these 
strings as the fivebrane is dragged across the threebranes.

As has been pointed out by various people \cite{wittenbaryon,groguri}, in the 
context of the anti-de~Sitter/conformal field theory ($AdS$/CFT)
correspondence there are general reasons to expect $N$ 
fundamental strings to join together on a D5-brane wrapped on a
five-sphere in the throat region of the threebrane geometry (i.e. in the $AdS$ 
geometry). This is the string theory counterpart to the gauge theory 
$SU(N)$ baryon vertex, representing a bound state of $N$ external quarks.
The baryon vertex has been studied in \cite{baryonsugra}, where the 
strings and the fivebrane are 
described in terms of separate Nambu-Goto actions. 
That approach ignores the worldvolume gauge field on the fivebrane. 
Its inclusion leads to the Born-Infeld action, which 
allows a unified description of the fivebrane and the strings. When 
restricted to the $AdS$ background, our solutions provide an explicit 
string theory representation of the baryon vertex. 
The Born-Infeld action for the worldbrane dynamics of the 
fivebrane in a threebrane background is an accessible and instructive way
to go after the energetics of this problem.

\subsection{The Setup}
\label{secSetup}

We set up the equations for the Born-Infeld D5-brane in the background 
geometry of a stack of $N$ D3-branes. The metric in a standard 
coordinate system is
$$
ds^2=H(r)^{-1/2}(-dt^2+dx_{||}^2)+H(r)^{1/2}(dr^2+r^2 d\Omega_5^2), \qquad 
H(r)=a+R^4/r^4~.
$$
We have chosen to express $H(r)$ in terms of an auxiliary constant 
$a$, in order to treat
the asymptotically flat D3-brane ($a=1$)
and the $AdS_5\times {\rm \bf S}^{5}$ ($a=0$) geometries in parallel.
The worldvolume action is the Born-Infeld action calculated using the 
induced metric (including the worldvolume gauge field)
$$
g^{ind}_{\alpha\beta} = g_{MN}\partial_\alpha X^M \partial_\beta X^N + 
{\cal F}_{\alpha\beta},
$$
plus the WZW term induced by the five-form field strength; the
latter is basically a 
source term for the worldvolume gauge field. The explicit 
action we will use is
$$
S = -T_5\int d^6\xi\sqrt{-det(g^{ind})} +T_5 \int d^6\xi A_\alpha
        \partial_\beta X^{M_1}\wedge\ldots\partial_\gamma X^{M_5}G_{M_1\ldots 
M_5}~,
$$
where $T_5$ is the D5-brane tension and the second term is the explicit
WZW coupling of the worldvolume gauge field $A$ to the background 
five-form field strength $G$. We use the target space time and 
${\rm \bf S}^{5}$ spherical coordinates as worldvolume coordinates for the 
fivebrane, $\xi_{\alpha}=(t,\theta_{\alpha})$. 

We pick a five-sphere surrounding a point on the threebrane stack and look 
for static solutions of the form $r(\theta)$ and $A_0(\theta)$ (with all 
other fields set to zero), where $\theta$ is the polar angle in spherical 
coordinates. Non-static solutions are of interest too, but we will not
deal with them in this paper. 
On substituting explicit forms for the threebrane metric and the five-form 
field strength, the action (for static configurations) simplifies to
\be
\label{action}
S= T_5 \Omega_{4}\int dt d\theta \sin^4\theta \{ -r^4H(r)
  \sqrt{r^2+(r^\prime)^2 -F_{0\theta}^2}  +4 A_0 R^4 \},
\ee
where $\Omega_{4}=8\pi^{2}/3$ denotes the volume of the unit four-sphere.
 
The gauge field equation of motion following from this action reads 
$$
   (\sin\theta)^{-4} \partial_\theta \left[-\sin^4\theta 
{(a r^4+R^4)E\over\sqrt{r^2+{r^\prime}^2-E^2}}\right] =  4 R^4~,
$$
where we have set $E=F_{0\theta}$ and the right-hand side is the 
source term coming from the WZW
action. It is helpful to repackage this in terms of the displacement $D$ (the 
variation of the action with respect to $E$):
\be
\label{dispeq}
D={\sin^4\theta (a r^4+R^4)E\over \sqrt{r^2+{r^\prime}^2-E^2}}
\quad \Rightarrow \quad
     \partial_\theta D(\theta) = -4 R^4 \sin^4\theta.
\ee

Obviously, we can integrate the equation for $D$ to find it as an
explicit function of $\theta$. The result is
\be
\label{d}
D(\theta) = R^4\left[{3\over 2}(\nu\pi-\theta) 
  +{3\over 2}\sin\theta\cos\theta+\sin^{3}\theta\cos\theta\right],
\ee
where the integration constant has been written in terms of a parameter 
$0\le\nu\le1$, whose meaning will be elucidated below. 
Notice that the sign of 
the WZW term in (\ref{action}) reflects the choice of a particular 
fivebrane orientation. Choosing the opposite orientation 
therefore reverses the sign of the source term in (\ref{dispeq}), and 
consequently the sign of $D$.

Since $D$, unlike $E$, is completely unaffected by 
the form of the function $r(\theta)$, it makes sense to express the 
action in terms of $D$ and regard the result as a functional for
$r(\theta)$. It is best to do this by a Legendre transformation, 
rewriting the original Lagrangian as
$$
U =  T_5 \Omega_{4}\int d\theta\{D\cdot E+ \sin^4\theta (a r^4+R^4)
    \sqrt{r^2+(r^\prime)^2-E^2}\}~.  
$$
Integrating the $DE$ term by parts using $E=-\partial_\theta A_0$, 
one reproduces (with a sign switch) the original Lagrangian. 
Using (\ref{dispeq}) we can eliminate $E$ in favor of $D$ to get
the desired functional of $r(\theta)$ alone:
\be
\label{u}
U =  T_5 \Omega_{4}\int d\theta 
\sqrt{r^2+(r^\prime)^2}\sqrt{D^2+(a r^4+R^4)^2\sin^8\theta}.
\ee

This functional is reasonably simple, but complicated by
the fact that there is explicit dependence on $\theta$. Hence there
is no simple energy-conservation first integral that we can use to
solve the equations (or at least analyse possible solutions).
For future reference, we record the Euler-Lagrange equations that
follow from (\ref{u}):
\bea
\label{bigel}
{d\over d\theta}\Bigl({r^\prime\over\sqrt{r^2+{r^\prime}^2}}
     \sqrt{D^2+(a r^4+R^4)^2\sin^8{\theta}} \Bigr) =
 {r\over\sqrt{r^2+{r^\prime}^2}} \sqrt{D^2+(a r^4+R^4)^2\sin^8{\theta}}
      \nonumber \\
\hspace{4cm}
  +{\sqrt{r^2+{r^\prime}^2}\over r}{4 a r^4(a r^4+R^4)\sin^8\theta\over
\sqrt{D^2+(a r^4+R^4)^2\sin^8\theta}}~.
\eea
Supersymmetry considerations will allow us to go rather far in analysing the
solutions of this formidable-looking equation.

When we discuss solutions in more detail, we will see that
it will not be possible to wrap the fivebrane smoothly around a sphere. 
Even if $r(\theta)\sim r_0$ for most $\theta$, we will find that for 
$\theta\to \pi$ (or 0), $r$  shoots off to infinity in a way that simulates a 
bundle of fundamental strings in 
the manner described in \cite{cm,gibbons}. 
Using (\ref{u})
we can already verify that the energy of such a spike is consistent with 
its interpretation as a bundle of strings. Suppose that the spike sticks out 
at $\theta=\pi$; then $D$ will take on some finite value $D(\pi)$ at
$\theta=\pi$. 
As we go into the spike, $r^\prime$ will dominate $r$ and the $D$ 
term will dominate $\sin^8\theta$. It is clear then that the spike has a 
`tension' (i.e. an energy per unit radial coordinate distance) 
$T_5 \Omega_{4} |D(\pi)|$.
Using the known facts that $D(\pi)=3\pi (\nu-1) R^4/2$ and 
$T_{5}\Omega_{4}R^{4}=2 N T_{f}/3\pi$, 
it follows that the `tension' of the spike is that of $n$ fundamental 
strings, $n T_f$, where $n=(1-\nu)N$. This gives meaning to the 
parameter $\nu$. 

\subsection{Supersymmetry Issues}
\label{secSUSY}

We are interested in placing a D5-brane in a D3-brane background and finding a 
structure
that looks like fundamental strings attached to the D3-branes. Insight into
what 
is possible 
is often obtained by looking for brane orientations such that the various
brane 
supersymmetry
conditions are mutually compatible for some number of supersymmetries. In type 
IIB
supergravity in ten-dimensional flat space, there are 32 supersymmetries 
generated by two
16-component constant Majorana-Weyl spinors $\eta_L,\eta_R$ of like parity
($\Gamma_{11}\eta_{L,R}=\eta_{L,R}$). In the presence of branes of various 
kinds, the 
number of supersymmetries is reduced by the imposition of further conditions. 
Explicitly,
\bea
\label{joe} \nonumber
{\rm F-string\ :} & \Gamma^{09}\eta_L=-\eta_L, & \Gamma^{09}\eta_R=+\eta_R, \\ 
{\rm D3-brane:} & \Gamma^{0123}\eta_L=+\eta_R, & \Gamma^{0123}\eta_R=-\eta_L,
\\ 
\nonumber
{\rm D5-brane:} & \Gamma^{045678}\eta_L=+\eta_R, & 
\Gamma^{045678}\eta_R=+\eta_L,
\eea
where the particular gamma matrix products are determined
by the embedding of the relevant branes into ten-dimensional space. For 
instance, the
D3-brane condition refers to a brane that spans the 123 coordinate directions. 
Conditions
can be multiplied by an overall sign by changing brane 
orientation.\footnote{Notice, however, that to maintain a 
supersymmetric configuration one must simultaneously reverse the orientation 
of two of the three objects.}
The existence of a BPS
state containing more than one type of brane depends on the existence of 
simultaneous
solutions of more than one of the above equations. The relevant point for our 
discussion is
that the conditions precisely as written above, corresponding to mutually 
perpendicular
D3-branes, D5-branes and F-strings, are compatible with eight supersymmetries 
(${\cal N}=2$ in 
usual parlance). The supersymmetry argument suggests that mutually orthogonal 
branes 
spanning a total of eight dimensions joined by a fundamental string running 
along the one
remaining dimension (perpendicular to both branes) should in fact be a stable 
BPS state.
An interesting aspect of our Born-Infeld worldvolume approach is that we will
explicitly see how the fundamental strings are created and destroyed as the 
D-branes are moved
past each other in the ninth direction (the Hanany-Witten effect 
\cite{hw, bdg, dfk}).

The above analysis has been carried out in flat space. To make contact with
the 
$AdS$/CFT correspondence, one would want to consider $N$ superposed D3-branes 
with $N$ large, in which case the background geometry is not flat and the 
supersymmetry analysis given above is at least incomplete. 
Imamura \cite{Imamura} has analysed the supersymmetry conditions 
associated with a D5-brane stretched over some surface in the `throat' of 
the D3-brane (where the geometry is
$AdS_5\times {\rm \bf S}^{5}$ and there is a flux of the RR five-form 
field strength
through the ${\rm \bf S}^{5}$). There are several new features here: first,
the 
unbroken supersymmetries of type IIB supergravity in this particular 
background are 32 position-dependent spinors (as opposed
to constant spinors in flat space); second, because of the RR five-form
field strength, there is a worldbrane gauge field induced on the 
worldvolume of the D5-brane; third, the condition that a particular element 
of the D5-brane worldvolume preserve some supersymmetry
involves the local orientation of the brane, the value of the induced 
worldvolume gauge field and the local value of the supergravity 
supersymmetry spinors. Since the D5-brane is embedded in
some nontrivial way in the  geometry, the supersymmetry condition is in 
principle different at each point on the worldvolume and it is far 
from obvious that it can be satisfied everywhere. 

However, Imamura \cite{Imamura} was able to show that these conditions boil
down, at least in the $AdS_5$ ($a=0$)
background, to a first-order equation for the embedding of the D5-brane into
the 
space
transverse to the D3-brane stack. In our language, his BPS condition can be 
written 
\be
\label{adsbps}
{r^\prime\over{r}} = {R^4\sin^5\theta +D(\theta)\cos\theta 
\over
     {R^4\sin^4\theta\cos\theta -D(\theta)\sin\theta} }~,
\ee
where $r=r(\theta)$ is the D5-brane embedding in the transverse space and 
$D(\theta)$ is the `displacement' field describing how the worldvolume 
gauge field varies from point to point. It is easy to show that any 
function $r(\theta)$ that satisfies this condition
automatically satisfies the Euler-Lagrange equation (\ref{bigel}) with
$a=0$; in that sense it is a first integral of the usual second-order equations.
Note that, as mentioned above, the structure of the action is such that there 
is no trivial energy first integral. The BPS condition (\ref{adsbps}) can be 
integrated analytically to obtain a two-parameter family of curves that 
describe BPS 
embeddings of a D5-brane into the $AdS_5\times {\bf S}^{5}$ geometry. 
These solutions 
will be discussed in the next section. 

We are also interested in exploring the analogous solutions in the
full asymptotically flat D3-brane 
background ($a=1$). In this background the 
interpretation and energetics of solutions should be quite straightforward. 
What is less obvious is how to find
BPS solutions. To follow Imamura's approach, one would first find the 
supersymmetry spinors in
the D3-brane background, use them to construct local supersymmetry conditions 
for an 
embedded D5-brane and from this find the condition on the embedding for there
to be a global worldvolume supersymmetry. 
This is no doubt perfectly feasible but we have not had
the patience to try it. Instead, we have simply guessed a generalization of
the $AdS_5\times{\bf S}^{5}$
BPS condition that automatically provides a solution of the Euler-Lagrange 
equations in
the full D3-brane background. The generalized BPS condition is obtained by 
making the (very plausible) replacement $R^4\to R^4+r^4$ in (\ref{adsbps}), 
\be
\label{d3bps}
{r^\prime\over{r}} = {(R^4+r^4)\sin^5\theta +D(\theta)\cos\theta 
\over
     {(R^4+r^4)\sin^4\theta\cos\theta -D(\theta)\sin\theta} }~.
\ee
It is easy to verify, using only (\ref{dispeq}), that this equation implies
the full Euler-Lagrange equation (\ref{bigel}) with $a=1$, so it is
certainly a first integral. Given its origin, it is almost certainly
the BPS condition as well. It is rather surprising that things work 
so smoothly, and we take this as another evidence of the special nature 
of the D3-brane background. The first-order equation (\ref{d3bps})
must be integrated numerically (as far as we know) and yields a two-parameter 
family of solutions whose structure is quite non-trivial. Exploration of these 
and the $AdS$ solutions will be the subject of the rest of the paper.

Before closing this section we note that in either background
one can obtain an `alternative' BPS condition by reversing the signs in 
front of $D$ in equations (\ref{adsbps}) or (\ref{d3bps}). 
The resulting condition would guarantee the preservation of a 
different set of supersymmetries, albeit just as many. In order to 
still have a first integral of the Euler-Lagrange equation (\ref{bigel}), 
$D$ must satisfy (\ref{dispeq}) with the opposite sign for the source term. 
Such oppositely oriented fivebrane configurations will actually have
the same embedding $r(\theta)$. 

\subsection{$AdS$ background: Born-Infeld Baryons}

\noindent
We start with a discussion of the solutions of the $AdS$ BPS equation 
(\ref{adsbps}) for the supersymmetric embedding of a fivebrane 
in the $AdS_{5}\times {\rm \bf S}^{5}$ geometry, with topology 
${\rm \bf S}^{4}\times {\rm \bf R}$. Fortunately, the BPS equation 
has the following simple analytic solution:\footnote{We thank 
\O.~Tafjord for help in finding this solution.}
\be
\label{adssol}
r(\theta)=\frac{A}{\sin\theta}
          \left[\frac{\eta(\theta)}{\pi(1-\nu)}\right]^{1/3},
\qquad \eta(\theta)=\theta-\pi\nu-\sin\theta \cos\theta,
\ee
where the scale factor $A$ is arbitrary, and $\nu$ is the integration
constant in (\ref{d}). The freedom of changing 
$A$ is a direct consequence of the scale invariance 
of the $AdS$ background: if $r(\theta)$ is a solution of 
(\ref{adsbps}), then so is $\lambda r(\theta)$ for any $\lambda$. 
Note that $\eta>0$ (so that the solution makes sense) only for 
$\theta_{cr}<\theta<\pi$, where $\theta_{cr}$ is defined by 
\be
\label{nu}
\pi\nu=\theta_{cr}-\sin\theta_{cr}\cos\theta_{cr}. 
\ee
This critical angle increases monotonically from zero to $\pi$
as $\nu$ increases from zero to one. Furthermore, when $\theta_{cr}>0$,
$r(\theta_{cr})=0$, a fact whose consequences will be explored 
below.
 
The fact that (if $\nu\neq 1$) the solution diverges as
$r\sim A/(\pi-\theta)$ when $\theta\to\pi$ 
means that a polar plot of $r(\theta)$ has, asymptotically, 
the shape of a `tube' of radius $A$. (This way of describing the
surface is a bit misleading as to the intrinsic geometry, but 
helps in visualization.) This tube is to be interpreted as a
bundle of fundamental strings running off to infinity in the space 
transverse to the D3-branes.
As explained in subsection \ref{secSetup}, the asymptotic `tension' of the tube 
equals that of $(1-\nu)N$ fundamental strings. For the classical 
solutions $\nu$ is a continuous parameter, but at the quantum 
level $\nu$ should obey the quantization rule $\nu=n/N$.
  
\begin{figure}             
 \begin{center}
 \leavevmode
 \epsfbox{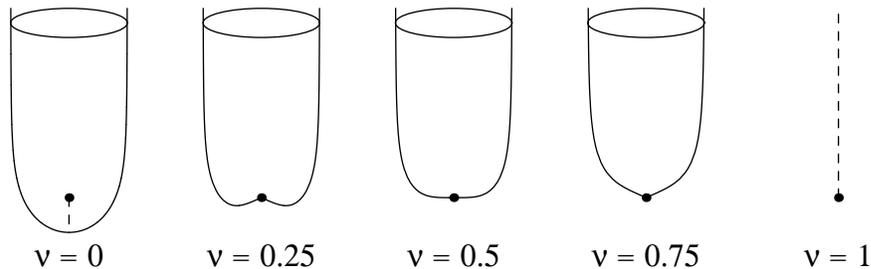}
 \end{center}
\caption[]{Polar plots of $r(\theta)$ for $AdS$ `tube' solutions 
   corresponding to $(1-\nu)N$ strings (with $\theta=\pi$ at the
   top of the plots). A 
  cross-section of each `tube' is an ${\rm \bf S}^{4}$.}
\label{adstubes}    
\end{figure}

In Figure.~\ref{adstubes} we have plotted (\ref{adssol}) for some representative values
of $\nu$. Consider first the special case $\nu=0$, which yields a tube 
with the maximal asymptotic tension $N T_{f}$ and corresponds to the
classic `baryon' vertex. In this case the solution starts at a finite 
radius $r(0)=(2/3\pi)^{1/3}A$, with $r'(0)=0$, and then
$r(\theta)$ increases monotonically with $\theta$. 
The initial radius $r(0)$ represents another way of setting the overall 
scale of this scale-invariant solution. 
The fact that the fivebrane surface stays away from
the horizon at $r=0$ suggests that it is well-decoupled from degrees
of freedom living on the the threebranes.
As far as the BPS equation is concerned, it seems to make sense to
consider the $\nu>0$ solutions as well. They have instructive features, 
although we will eventually conclude that they are on a less sound 
footing than their $\nu=0$ cousins. For large $r$, the solution asymptotes 
to the familiar tube with a tension corresponding to $(1-\nu)N<N$ strings:
it corresponds to a general multi-quark state of a $U(N)$ gauge theory. 
As mentioned above, for $\nu> 0$, the surface intersects $r=0$  
at an angle $\theta_{cr}>0$ defined by (\ref{nu}), leading to the
cusp-like configurations displayed in Fig.~\ref{adstubes}. Note that, because 
the $r\to 0$ cusp has a finite opening angle, the fivebrane
does not capture all of the five-form flux: this is closely related
to the fact that the asymptotic tension is $(1-\nu)N$ and not $N$.

As $\nu\to 0$, the opening angle $\theta_{cr}\to 0$. The approach 
to the $\nu=0$ solution, which does not contact $r=0$, is achieved, 
as shown in Fig.~\ref{adstubes}, via a `tensionless string' connecting the minimum
radius of the $\nu=0$ solution to $r=0$ (indicated as a dotted 
line in the figure). At the other extreme, $\nu\to 1$, one has 
$\theta_{cr}\to\pi$, and the solution collapses to a similar 
`phantom string', this time running from $r=0$ to infinity.

One can compare the total energy of these configurations to that of 
$(1-\nu)N$ fundamental strings (this was done in \cite{Imamura} for 
the case of $\nu=0$). Using the solution (\ref{adssol}) in expression
(\ref{u}), the energy of the fivebrane up to an angular cutoff 
$\theta_{max}$ can be put in the form
\be
\label{uconfig}
U(\theta_{max})=NT_{f}\,\frac{A}{\pi}\int^{\theta_{max}}_{\theta_{cr}}
   d\theta\,\left[\frac{\eta(\theta)}{\pi(1-\nu)}\right]^{1/3}
   \left\{\frac{\eta(\theta)}{\sin^{2}\theta}-\frac{4}{3}\sin\theta
   \cos\theta+\frac{4\sin^{4}\theta}{9\eta(\theta)}\right\}.
\ee
The fundamental string energy, on the other hand, for strings 
extending from the origin to a radial cutoff $r_{max}=r(\theta_{max})$,
is simply 
$E_{str}(\theta_{max})=(1-\nu)N T_{f}\,r(\theta_{max})$. 
It is easy to check numerically that  
$E_{str}(\theta_{max})-U(\theta_{max})\to 0$ as 
$\theta_{max}\to\pi$ ($r_{max}\to\infty$). 
The Born-Infeld fivebrane `tubes' can be therefore 
regarded as threshold bound states 
of $(1-\nu)N$ fundamental strings. We emphasize that this holds for any 
value of the scale parameter $A$: as $\theta_{max}\to\pi$, the energy 
$U(\theta_{max})$ becomes independent of $A$. The parameter $A$ is therefore 
a modulus in the space of equal-energy solutions. 

A complication for the interpretation of these solutions is that, in general
(specifically, when $\nu\ne 0,1/2,1$), the total five-form flux captured 
by the fivebrane differs from the number of fundamental strings, $(1-\nu)N$,
indicated by the asymptotic tension or total energy. The fundamental string 
charge is the source of the displacement field $D$, and we can rearrange 
(\ref{dispeq}) to show that a fivebrane that runs from $\theta=\theta_{cr}$
to $\theta=\pi$ intercepts a total five-form flux
$$
Q_{flux}= -{2N\over 3\pi R^4}\left[ D(\pi)-D(\theta_{cr})\right]
         = (1-\nu)N + {2N\over 3\pi} \sin^3\theta_{cr} 
         \cos\theta_{cr} ~.
$$
{}From the value of the tension, we would have expected a total 
charge $Q_{tot}=(1-\nu)N$ on the D5-brane. The difference,\footnote{
It should be clear that $Q_{flux}$, $Q_{tot}$, and 
$Q_{missing}$ all change sign if we reverse the fivebrane orientation.} 
\be
\label{qmiss}
Q_{missing}= -{2N\over 3\pi R^4} \sin^{3}\theta_{cr}\cos\theta_{cr}~,
\ee
is nonzero for $\nu\neq 0,1/2,1$ and presumably must be accounted for 
by a point charge at $r=0$. Since $r=0$ is where the fivebrane makes 
contact with the threebranes, this reminds us that, in order to be
fully consistent, we should take into account the possibility of exciting
the threebrane worldvolume $U(N)$ gauge fields when we attach fundamental
strings to the D3-branes (as in \cite{cm,gibbons}). The case of $N$ 
strings ($\nu=0$) is special since they can be in a $U(N)$ singlet which 
will decouple from the D3-brane worldvolume gauge theory. When $\nu\ne 0$,
we are talking about a collection of strings that cannot be $U(N)$ neutral
and must excite the D3 gauge fields, which will in turn react back on the
metric. Since we have not allowed for this possibility in our construction,
the detailed features of our solutions with $Q_{missing}\ne 0$ have to be 
taken with a certain grain of salt. The case $\nu=1/2$ is peculiar:
it corresponds to $N/2$ strings and so cannot form a $U(N)$ singlet, yet
has $Q_{missing}=0$. We are not sure that it really has the same
status as the true singlet $\nu=0$ solution. 

In light of the $AdS$/CFT correspondence \cite{jthroat,gkpads,wittenholo},
the above results are expected to have a gauge theory interpretation.
As discussed by several people \cite{wittenbaryon, groguri}, 
a baryon (a bound state of $N$ external quarks) in the $SU(N)$ ${\cal N}=4$ 
supersymmetric Yang-Mills (SYM) theory corresponds, 
in the dual $AdS$ description, to $N$ fundamental strings 
which join together on a D5-brane wrapped on an ${\rm \bf S}^{5}$ at some 
radius. The Born-Infeld $\nu=0$ fivebrane configuration described 
above provides a detailed representation of such a baryon. 
In particular, the absence of binding 
energy is as expected for a BPS threshold bound state in the
${\cal N}=4$ theory. Our other solutions with $\nu=n/N$ ($0<n<N$) 
are also BPS and correspond to threshold bound states of $N-n$ quarks. 
The existence of color non-neutral states with finite (renormalized) energy
is perfectly reasonable in a non-confining theory. To start learning
something interesting about these states, we would have to go beyond
mere energetics and ask some dynamical questions. 
To be absolutely clear, we emphasize that in every case discussed here
the quarks in the gauge theory are all at the same spatial location.

The solutions that we have discussed so far
are naturally restricted to the range of angles 
$\theta_{cr}\leq\theta\leq\pi$ where $\eta(\theta)>0$. 
We will call them `upper tubes'. A simple modification
of (\ref{adssol}) is valid for the complementary angular range 
$0\leq\theta\leq\theta_{cr}$ where $\eta(\theta)<0$:
\be
\label{adstildesol}
\tilde{r}(\theta)=\frac{\tilde{A}}{\sin\theta}
          \left[\frac{-\eta(\theta)}{\pi\nu}\right]^{1/3}.
\ee
This expression is singular at $\theta=0$
(where $\tilde{r}(\theta)\sim \tilde{A}/\theta$) and meets the origin at
$\theta=\theta_{cr}$. It represents a downward-pointing tube of 
`radius' $\tilde{A}$ whose shape and tension are the same as for 
an upward-pointing tube with parameter $1-\nu$. 
In other words, $\tilde{r}(\theta;\nu)=r(\pi-\theta;1-\nu)$.
This `lower tube' solution intercepts a total flux
$Q_{flux}=-\nu N + 2 N \sin^3\theta_{cr}\cos\theta_{cr}/3\pi$.
{}From the tension of this configuration, we would have expected a total
charge 
$Q_{tot}=-\nu N$, so there is a charge $Q_{missing}$ localized at the 
origin which is again given by (\ref{qmiss}). If the `upper tube' solution
corresponds to some number of quarks, the `lower tube' solution corresponds
to some number of antiquarks. 

\begin{figure}[ht]            
 \begin{center}
 \leavevmode
 \epsfbox{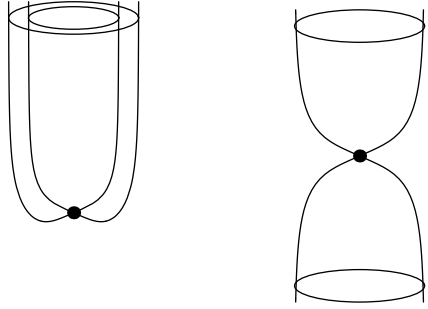}
 \end{center}
 \caption[]{Upper/upper and upper/lower tube combinations. These 
  configurations have vanishing charge at the origin (see text).}
 \label{tubecomb}      
\end{figure}

Finally we want to speculate about constructing new solutions
by combining the $\nu\ne 0$ solutions we have been discussing.
Specifically, we are interested in obtaining configurations for which 
the peculiar charge at the origin cancels. 
Inspection of (\ref{qmiss}) shows 
that this can be achieved by merging two tubes whose opening angles 
are complementary. Using equation (\ref{nu}) this means that if one of 
the tubes has parameter $\nu$, the other one must have
parameter $1-\nu$. 
Taking into account the possibility of using `upper' or `lower' solutions, 
one is thus led to two types of configurations, illustrated
in Fig.~\ref{tubecomb}. 

The combination of two upper tubes with parameters $\nu$ 
and $1-\nu$ yields a baryon-like configuration corresponding to 
a total of $N$ quarks. This system differs from the $\nu=0$ baryon 
of Fig.~\ref{adstubes} in that the `strings' have been separated into two 
distinct coaxial tubes. 
It is interesting to note that this combined structure can be 
obtained as a single solution of (\ref{adsbps}), 
with a unique value of $\nu$, by formally continuing $r(\theta)$ 
in (\ref{adssol}) beyond 
$\theta=\theta_{cr}$ (where $r=0$) to negative values of $r$.
The continued solution, depicted in Fig.~\ref{tubecomb},
can be interpreted as a single surface which
intersects itself at the origin.
In this interpretation the parameter $\nu$ is an additional modulus
of the baryon, controlling how many strings (out of the total of $N$)
`lie' in each tube.

If instead one puts together an upper $\nu$ and a lower $1-\nu$ 
solution, the result represents $1-\nu$ strings which 
extend from $r=\infty,\theta=0$ to $r=\infty,\theta=\pi$,
and run through the origin. 
In the gauge theory language this describes a state with 
$1-\nu$ quarks and the same number of antiquarks. 
(This is still BPS, because the quarks and antiquarks have 
opposite $SU(4)$ quantum numbers.) The total charge of the 
state vanishes.   

Judging from the cancellation of the charge at the origin, these 
combined solutions would appear to have the same status as the baryon. 
On the other hand, it is unclear to what extent these superposed 
tubes can be regarded as a single object, given that they are 
`connected' only at the infinitely distant point $r=0$. It 
would be interesting to see whether fluctuations propagating
inward along one tube can `tunnel' through the point at $r=0$ 
to propagate outward along the second tube.

\section{Hanany-Witten effect in the full D3-brane metric background}
\label{hw}

So far, we have looked at the static solutions of D5-branes in the
$AdS_5\times {\rm \bf S}^5$ geometry of the `throat' region of 
the exterior geometry of a large number of D3-branes. As we 
now know, this limit gives us a supergravity description of 
${\cal N}=4$ SYM theory. We can also shed light on
some old string theory questions by studying the same types of 
configurations in the full asymptotically flat geometry of multiple
D3-branes.
 
To examine the character of the solutions in the asymptotically flat 
D3-brane background, it is convenient to parametrize the solution by
$z=z(\rho)$, where $\rho=r\sin\theta$, and $z=-r\cos\theta$. In these 
variables, adapted to flat space, the BPS condition (\ref{d3bps}) reads
\be
\label{d3bpsz}
z^\prime(\rho)=\frac{-D(\arctan(-\rho/z))}{\rho^{4}
               \left(1+\frac{R^{4}}{(\rho^{2}+z^{2})^{2}}\right)}\, .
\ee
Solutions to this equation describe D5-brane configurations which asymptote 
to a flat plane as $\rho\to\infty$ (equivalently, $\theta\to\pi/2$). 
The leading asymptotic behavior following from (\ref{d3bpsz}) is
\be
\label{asymptotplane}
z(\rho)=z_{max}+\frac{D(\pi/2)}{3\rho^{3}}
        +{\mathcal{O}}\left(\frac{R^{4}}{\rho^{4}}\right), 
\ee
where $z_{max}$ denotes the transverse position of the flat brane. 
We will be interested in studying how the solution changes 
as we vary $z_{max}$.
 
\begin{figure}[ht]                  
 \begin{center}
 \leavevmode
 \epsfbox{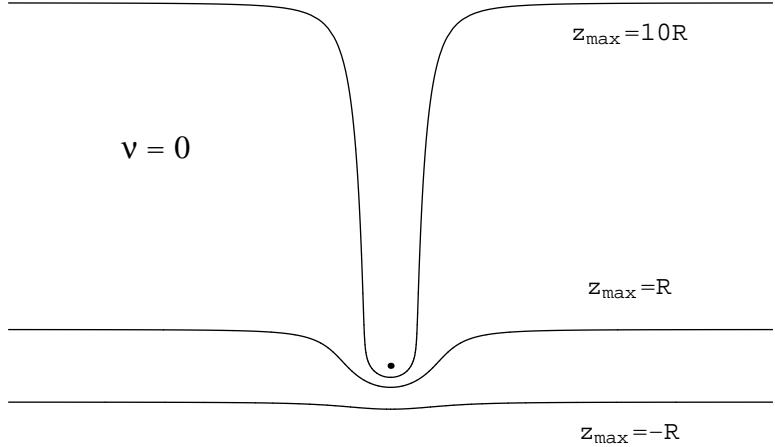}
 \end{center}
\caption[]{Solutions describing the creation of $N$ fundamental 
   strings as a D5-brane is dragged upward, across a stack of 
   D3-branes. The number of strings connecting the two types of branes changes
   from 0 to $N$.}
\label{d3nuzero}
\end{figure}

\begin{figure}[t]                  
 \begin{center}
 \leavevmode
 \epsfbox{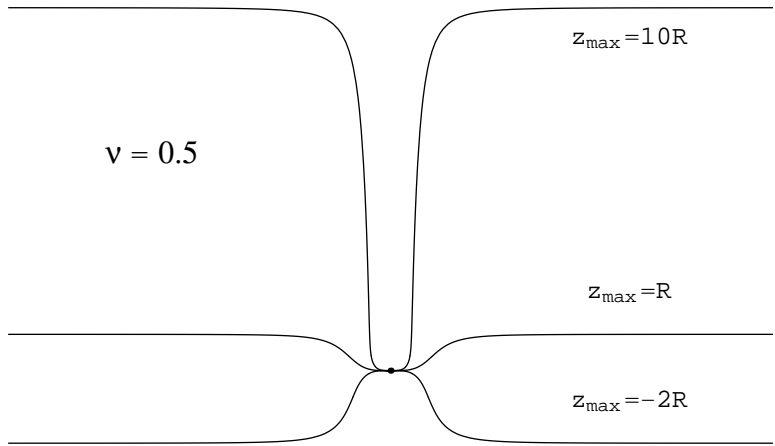}
 \end{center}
\caption[]{Solutions describing the creation of $N$ fundamental strings
   as a D5-brane is dragged across a stack of 
   D3-branes. The number of strings connecting the two types of branes changes
   from $-N/2$ to $+N/2$ (the signs indicate whether the strings
   originate or terminate on the fivebrane).}
\label{d3nuhalf}
\end{figure}

Figures \ref{d3nuzero} and \ref{d3nuhalf}
show the configurations obtained by integrating
(\ref{d3bpsz}) numerically for $\nu=0$ and $\nu=1/2$ and for a 
few representative values of $z_{max}$. The stack of $N$ D3-branes is at the 
origin, and extends along directions perpendicular to the figure.
For any value of $z_{max}$, the D5-brane captures the same fraction 
of the total five-form flux, which (in conjunction with a
possible point charge at the origin, as discussed in the previous
subsection) effectively endows the D5-brane with a total of 
$(1/2-\nu)N$ units of charge. Note the shift of $N/2$ units of charge, 
compared with the analogous situation in $AdS$ space: this happens 
because the asymptotic region of the brane is now at $\theta=\pi/2$
rather than $\theta=\pi$. This will have interesting consequences.

Consider first the situation for $\nu=0$, described graphically in 
Figure~\ref{d3nuzero}. 
As $z_{max}\to -\infty$, the charge density vanishes, and the D5-brane 
of course becomes flat. As one approaches the stack of threebranes 
from below ($z_{max}\to 0_{-}$), the charge becomes more and more 
localized near the center of the fivebrane, and the configuration becomes 
slightly deformed, bending away from the origin. 
As $z_{max}$ increases, the D5-brane remains `hung-up' on the 
D3-brane stack at $r=0$ and a tube of topology 
${\rm \bf S}^{4}\times {\rm \bf R}$ 
gets drawn out. The total charge of the tube itself approaches $N$
as it gets longer and longer and it becomes indistinguishable from a 
bundle of $N$ fundamental strings. Curiously, when the bundle eventually
connects to the flat D5-brane, a region of negative five-form flux is
encountered and the total charge intercepted by the fivebrane 
drops to $N/2$ (for any $z_{max}$). Altogether, then, this family 
of solutions provides a very concrete picture of the creation of 
fundamental strings as a fivebrane is dragged over a stack of 
threebranes, the Hanany-Witten effect \cite{hw, bdg, dfk}. 

For $\nu>0$ the story is modified in exactly the same way as in 
the previous subsection. For either sign of $z_{max}$, the fivebrane 
now reaches the origin, $r=0$, at an angle $\theta=\theta_{min}$ 
given in terms of $\nu$ by equation (\ref{nu}). 
As $z_{max}\to -\infty$ the solution describes now a fivebrane 
connected by $\nu N$ strings to the stack of threebranes. 
For definiteness, assume the choice of sign for $D$ (i.e., the 
orientation of the fivebrane) is such
that the strings emanate from the D5-brane and terminate on the D3-branes.
Upon moving past $z_{max}=0$, $N$ strings 
directed towards the fivebrane are created, 
and as $z_{max}\to \infty$ $(1-\nu)N$ strings directed 
away from the threebranes extend between the two types of branes. The 
case $\nu=1/2$ is portrayed in Figure~\ref{d3nuhalf}.

It is instructive to compare the solution described above to the 
description of fundamental strings attached to a fivebrane as
a Coulomb solution of the fivebrane Born-Infeld theory 
\cite{cm,gibbons}. In the latter case the parent brane is 
embedded in flat space and the worldvolume gauge field is simply 
that of a point charge. For $n$ units of charge, the spike configuration
that protrudes from the brane at the location of the charge is of the form  
\be 
\label{cm}
z(\rho)=z_{max}-\frac{n c_{5}}{\rho^{3}},
\ee
where $c_{5}=2\pi^{2}g_{s}(\ap)^{2}$ is the quantum of charge. 
Writing this in terms of the threebrane throat radius
$R=(4\pi N g_{s}(\ap)^{2})^{1/4}$, one can readily verify that the 
asymptotic form (\ref{asymptotplane}) agrees with the solution
(\ref{cm}) for $n=(1/2-\nu)N$ strings. This is precisely as one would expect 
from the above discussion, for the entire fivebrane captures precisely 
a fraction $(1/2-\nu)$ of the total five-form flux. 
By the same token, it is clear 
that the present solution is of a more complex nature than that of 
\cite{cm,gibbons}. The configuration discussed 
here corresponds roughly to a solution which is locally of the type 
(\ref{cm}), but where the charge $n$ varies as one changes position on 
the fivebrane.

One of the more confusing features of the Hanany-Witten effect is its
energetics: does the created string exert a force and, if so, how
is that consistent with the BPS property? We can shed some light on this by 
computing the energy of our configurations. In 
terms of the $z(\rho)$ parametrization, and using the fact that 
$T_{5}\Omega_{4}R^{4}=2 N T_{f}/3\pi$, equation (\ref{u}) becomes
\be
\label{uz}
U=N T_{f}\,\frac{2}{3\pi}\int d\rho\, 
\sqrt{1+\left(\partial_{\rho}z\right)^{2}}
\sqrt{D^{2}+\rho^{8}\left[1+\frac{R^{4}}{(\rho^{2}+z^{2})^{2}}\right]^{2}}\,.
\ee
We can use the BPS condition (\ref{d3bpsz}) to 
express the energy integrand
for our solutions solely as a function of $z$ and $\rho$,
\be
\label{uzrho}
U=N T_{f}\,\frac{2}{3\pi}\int d\rho\, 
 \left\{ \frac{D^2}{\rho^4\left[1+\frac{R^4}{(\rho^2+z^2)^2}\right]}
  +\rho^{4}\left[1+\frac{R^{4}}{(\rho^{2}+z^{2})^{2}}\right]\right\}.
\ee 

Now, the energy of the infinite D5-brane is evidently divergent, so 
(\ref{uzrho}) must be regularized. If we do so by placing a cutoff 
$\rho_{c}$ on $\rho$, the leading and subleading contributions
to (\ref{uzrho}) are clearly quintic and linear in $\rho_{c}$, 
respectively. 
The offending terms, however, are independent of 
$z_{max}$, so we choose to simply drop them (thereby removing an 
infinite constant from $U$). Altogether, then, we subtract 
$\rho^{4}+R^{4}$ from the integrand of (\ref{uzrho}) to obtain the
expression
\be
\label{usub}
\hat{U}(z_{max},\rho_{c})=N T_{f}\,\frac{2}{3\pi}\int^{\rho_{c}}_{0}d\rho\, 
 \left\{\frac{D^2}{\rho^4\left[1+\frac{R^4}{(\rho^2+z^2)^2}\right]}
  -\frac{2\rho^{2}z^{2}+z^{4}}{(\rho^{2}+z^{2})^{2}}\right\},
\ee 
which is finite as the cutoff is removed. Using the numerical solutions, 
one finds that in fact
$\hat{U}\to (1/2-\nu) N T_{f}\, z_{max}$ 
for all $z_{max}$ as $\rho_{c}\to\infty$.

Since the difference between $U$ and $\hat{U}$ is a constant, it 
readily follows that there is a net constant force on the fivebrane,
independent of $z_{max}$:
\be
\label{force}
\frac{\partial}{\partial z_{max}}U(z_{max},\rho_{c}) \to
       \left(\frac{1}{2}-\nu\right) N T_{f}
\quad\mbox{as}\quad \rho_{c}\to\infty.
\ee
The total force equals the tension of $(1/2-\nu)N$ 
fundamental strings as a consequence of the fact that the full
configuration carries a total of $(1/2-\nu)N$ units of charge. It might seem 
surprising at first that there is a force on the fivebrane 
even for $\nu=0$ and $z_{max}<0$ 
(i.e., before the D5-brane crosses the 
D3-brane stack and a tube of strings is created),
but one must keep in mind that even then there is 
a charge on the brane, and consequently a position-dependent energy.
After the fivebrane is moved past the threebranes, to $z_{max}>0$, a 
bundle of fundamental strings is created, and this bundle pulls down 
on the fivebrane with a force which tends to $N T_{f}$ as 
$z_{max}\to\infty$. The net force on the brane is still $(1/2-\nu)N T_{f}$, 
however, because the outer portion of the brane now carries negative 
charge, as a consequence of which there is an additional, upward force 
on the brane.
Our approach thus makes it absolutely clear that, contrary to the naive
expectation, there is no discontinuity in the force as the branes cross.
One is able to find static solutions despite the 
presence of a constant force because the D5-brane is infinitely 
massive, and therefore will not move.
If desired, this constant force can be cancelled by placing 
$(1-2\nu)N$ additional D3-branes at $z=-\infty$.

An alternative way to reach the same conclusions is to compute the 
force by cutting off (\ref{uz}) at $\rho_{c}$ and differentiating with 
respect to $z_{max}$ under the integral (regarding 
$z=z(\rho;z_{max})$). After an integration by 
parts and an application of the Euler-Lagrange equation, one is left 
only with a boundary term, which yields the analytic expression
\be
\label{slope}
\frac{\partial U}{\partial z_{max}}(z_{max},\rho_{c})=
N T_{f}\,\frac{2}{3\pi}\left.\left\{ 
\frac{\partial_{\rho}z}{\sqrt{1+(\partial_{\rho}z)^{2}}}
\sqrt{D^{2}+\rho^{8}\left[1+\frac{R^{4}}{(\rho^{2}+z^{2})^{2}}\right]^{2}}
\frac{\partial z}{\partial z_{max}}\right\}\right|_{\rho_{c}}.
\ee
Using (\ref{slope}) and (\ref{asymptotplane}) one can again conclude 
that the force on the fivebrane approaches $(1/2-\nu)N T_{f}$ as 
$\rho_{c}\to\infty$ at a fixed $z_{max}$. Furthermore, 
using the numerical solution in conjunction with (\ref{slope}), one 
can compute the force on the $\rho<\rho_{c}$ portion of the brane, 
for any value of $\rho_{c}$. For any fixed $\rho_{c}$, it is easy to 
see that the force tends to $(1-\nu)N T_{f}$ ($\nu N T_{f}$)
as $z_{max}\to\infty$ ($z_{max}\to-\infty$). Taking the limit this
way picks out the stress on the string tube part of the configuration
and yields the expected tension of $(1-\nu)N$ ($\nu N$) fundamental 
strings.
Nonetheless, the total asymptotic stress on the D5-brane is smaller 
by $N/2$ fundamental string units, due to the extra five-form 
flux intercepted by the flat part of the D5-brane. 

Notice that for $\nu=1/2$ the net force on the D5-brane vanishes. This is 
because in that case the total charge on the 
brane is zero. As a result, the $\rho^{-3}$ term in 
(\ref{asymptotplane}) has a vanishing coefficient.
This configuration has $z( \rho=0) = z'(\rho=0)=0$,
$\theta_{min}= \pi/2$, and $D(\theta_{min}) =0$. It is only for this
and the $\nu=0$ and $\nu=1$ cases that
the point charge at the origin vanishes.
This solution (depicted in Fig.~3.4)
describes a bundle of $N/2$ strings which flip their orientation
as the fivebrane to which they are attached is moved 
above or below the threebrane stack.
The number of attached strings still changes by $N$,
from $-N/2$ to $+N/2$, as the D5-brane is pulled through the
stack.  This configuration is thus a realization 
 of the `half-string' ground-state of the system described in 
\cite{dfk}. For reasons explained in that paper, and confirmed
by our energy analysis, this is the only solution which is in a
state of neutral equilibrium.

\begin{figure}[ht]                
 \begin{center}
 \leavevmode
 \epsfbox{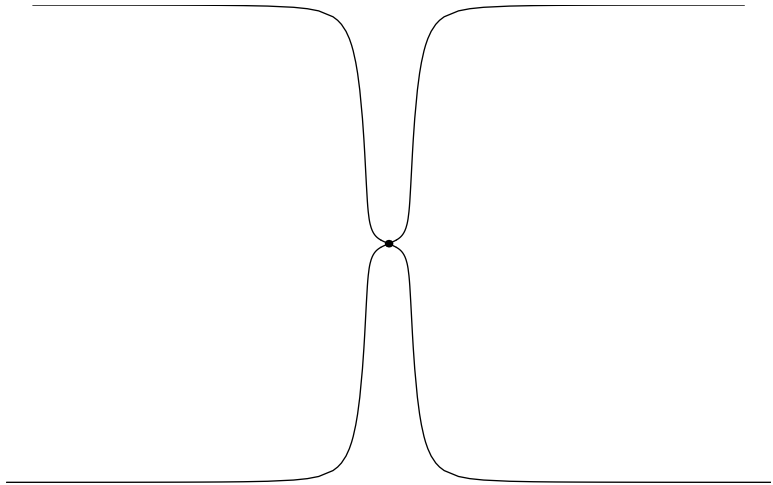}
 \end{center}
\caption[]{Solution describing a system of two parallel D5-branes
   connected by $(1-\nu)N$ fundamental strings which run through the 
   D3-branes at $r=0$. A `point W-boson charge' lies at the origin.}
\label{d5d5strings}  
\end{figure}

Just as in the previous subsection, one could imagine combining solutions
to obtain configurations in which the charge at the origin vanishes.
We will focus attention here on the possibility of superposing 
two solutions with parameters 
$\nu$, $z_{max}>0$ and $1-\nu$,$-z_{max}$, respectively. 
The complete structure obtained
this way is illustrated in Figure~{d5d5strings}, and corresponds
to a configuration in which two infinite parallel D5-branes with the same 
orientation, located at $\pm z_{max}$, are connected by $(1-\nu)N$ 
fundamental strings running through the $N$ D3-branes at the origin. 
Something interesting has happened here: we have constructed an
excitation of a system of two parallel fivebranes, something which
should more properly be described by the non-abelian $SU(2)$
Born-Infeld action. We have achieved this effect (perhaps illegitimately!)
by gluing together two $U(1)$ solutions at the singularity provided
by the D3-branes. The logic of this construction is similar in spirit
to a previous attempt to construct Prasad-Sommerfield monopole 
configurations from Born-Infeld dynamics \cite{aki}. Following that 
interpretation, the point charges of the component
solutions at the origin should be understood not to cancel, but 
to combine instead into a `point W-boson charge' which interpolates,
at no cost in energy, between the two $U(1)$s of the overall broken
$U(2)$ symmetry. 

The threebrane background geometry evidently plays a role
in facilitating the construction described above. Nevertheless,
since the `strings' in the solution are interpreted as 
merely passing through  
the D3-branes, it is natural to conjecture that there exist
neighboring static fivebrane configurations in 
which the strings miss the origin. Deforming the system in this manner 
it would be possible to move the connecting strings arbitrarily far 
away from the threebranes, thereby producing the analogous flat 
space configuration. It would be very interesting to pursue this issue 
further.

\section{The Baryon in Three Dimensions} \label{3dsec}

The Born-Infeld action plus an appropriate Wess-Zumino term defines
a worldvolume theory for D-branes which has proved to be a
powerful way to describe these objects and their excitations.
In the context of Maldacena's correspondence between supergravity
in anti-de Sitter ($AdS$) space and certain gauge theories
\cite{jthroat,gkpads,wittenholo},
there are gauge theory questions which can be answered by the study
of branes extended in curved space. In particular, it was shown in a
general way that the large-$N_c$ dynamics of baryons can be related 
to the
behavior of D5-branes extended in $AdS$ space \cite{wittenbaryon, 
groguri}.
Concrete realizations of this possiblity in the context of the 
non-confining
${\mathcal N}=4$ supersymmetric gauge theory have been worked out
in \cite{cgs,imamura}
using the Born-Infeld approach for constructing strings out of 
D-branes
\cite{cm,gibbons}. In this paper, we will look at how these 
constructions
work in the more complicated spacetimes that correspond to various 
confining
gauge theories. We will examine confining forces by looking at what 
happens
to baryons when they are pulled apart into their quark constituents. 
This
will be compared to (and yields somewhat more information than) the 
study
of confining forces via simple strings that `hang' into the $AdS$ 
geometry
from a boundary Wilson loop \cite{rey,juan, bisy2, groguri}.

In this Section, we extend our previous work \cite{cgs} on D5-branes in
extremal background in three ways. First, we will study non-extremal
supergravity backgrounds, corresponding to gauge theories 
dimensionally
reduced in a way that breaks supersymmetry. Second, we allow the 
brane
configurations to have extension in the spacetime coordinates of the
gauge-theory instead of being localized at a point. This will allow 
us
to describe a baryon which is being `pulled apart' into quark 
constituents.

We start out in Section \ref{3dsec} by analysing baryons in a 
(2+1)-dimensional $SU(N)$ Yang-Mills theory which is obtained from 
(3+1)-dimensions by compactifying on a supersymmetry breaking circle.
As proposed in \cite{wittenthermo}, this gauge theory is dual to
a certain non-extremal D3-brane geometry\footnote{See \cite{klebanov,
minahan, others} 
for 
some interesting alternative string theory approaches 
to the study of large-$N$ non-supersymmetric Yang-Mills theories.} 
and, following \cite{cgs},
we study solutions of the D5-brane worldvolume equations of motion
in that geometry. We find a class of solutions that are localized in
the gauge theory spatial coordinates and appear to describe the
baryon. Unlike the baryons constructed in the extremal background
\cite{cgs,imamura}, these solutions have no moduli since the quarks
are truly bound in the baryon. We then study a new class of solutions
in which the $N$ quarks are separated into two groups, containing
$\nu N$ and $(1 - \nu)N$ quarks respectively, separated by a spatial
distance $L$ in the gauge theory. The $L$-dependence of the energy of
these solutions is consistent with confinement and the implied color
flux tube tension has a non-trivial dependence on the color charge 
$\nu$.

\subsection{Worldvolume Action and Equations of Motion}
\label{eqn3dsec}

We derive the equations for a D5-brane embedded in the near-horizon
geometry of $N$ nonextremal D3-branes. The Euclidean background metric is
\bea \label{d3metric}
{}&{}&ds^2=\left(\frac{r}{R}\right)^{2}\left[f(r)d\tau^2+dx_{||}^2\right]
+\left(\frac{R}{r}\right)^{2}f(r)^{-1}dr^2+R^2 d\Omega_5^2, \\
{}&{}&f(r)=1-r_{h}^4/r^4, \qquad R^{4}=4\pi\gs N \ls^{4}, \qquad
r_{h}=\pi R^{2} T, \nonumber
\eea
where $\{\tau,x_{||}\}=\{\tau,x,y,z\}$ denote the directions parallel to the
threebranes. The coordinate $\tau$ is periodic, with period
$1/T$, where $T$ is the Hawking temperature.
The relation between the horizon radius $r_{h}$ and $T$ ensures smoothness
of the geometry at $r=r_{h}$.

Under the $AdS$/CFT correspondence \cite{jthroat,gkpads,wittenholo},
type IIB string theory on a background with the above metric, a
constant dilaton, and $N$ units of fiveform flux through the five-sphere,
is dual to ${\cal N}=4$, $d=3+1$ $SU(N)$ SYM theory at temperature
$T$, with coupling $\g{4}^{2}=2\pi\gs$. The gauge theory coordinates
are $\{x,y,z,\tau\}$. For large $T$ the $\tau$
circle becomes small and one effectively obtains a description of a
strongly-coupled (coupling $\g{3}^{2}=\g{4}^{2}T$) three-dimensional
Euclidean gauge theory at zero temperature. The thermal boundary
conditions on the circle break supersymmetry and the fermions and
scalars acquire masses of order $T$ and $\g{4}^{2}T$, respectively.
The effective three-dimensional theory is expected to display behavior
similar to that of non-supersymmetric pure Yang-Mills theory,
$\mbox{QCD}_{3}$ \cite{wittenthermo}.

A baryon (a bound state of $N$ external quarks) in the three-dimensional
theory has as its string theory counterpart a fivebrane wrapped on an
${\bf S}^{5}$ on which $N$ fundamental strings terminate
\cite{wittenbaryon,groguri}. The fivebrane worldvolume action is
$$
S = -T_5 \int d^6\sigma\sqrt{\det(g+F)} +T_5 \int A_{(1)}\wedge G_{(5)}~,
$$
where $T_5=1/\left(\gs(2\pi)^{5}\ls^{6}\right)$ is the brane tension.
The Born-Infeld term involves
the induced metric $g$ and the $U(1)$ worldvolume
field strength $F_{(2)}=d A_{(1)}$.
The second term is the Wess-Zumino coupling of the
worldvolume gauge field $A_{(1)}$ to (the pullback of)
the background five-form field strength $G_{(5)}$, which effectively
endows the fivebrane with a $U(1)$ charge proportional to the
${\bf S}^{5}$ solid angle that it spans.

For a static baryon we need a configuration invariant under translations
in the gauge theory time direction, which we take to be $y$. 
We parametrize the world volume of the fivebrane by six parameters 
$\tau,~ \sigma_1,~\sigma_2,$$~ \sigma_3,~ \sigma_4$ and $s$. The embedding of this
fivebrane into 10-dimensional target space is defined by
\be 
\begin{array}{ll}
x=x(s),    &\theta_1=\sigma_1, \\
y=\tau,    &\theta_2=\sigma_2,  \\
z=Const=0, &\theta_3=\sigma_3,  \\
t=Const=0, &\theta_4 =\sigma_4, \\
r=r(s),    &\theta_5 = \theta(s),
\label{staticgauge}
\end{array} 
\ee
where only $x$, $\theta_5$ and $r$ are functions of the parameter $s$.
This restricts our attention to $SO(5)$ symmetric configurations, i.e.
the ones invariant under rotations about the principal polar axis 
with respect to which $\theta_5$ is measured.
The action then simplifies to
\be \label{d3action}
S= T_5 \Omega_{4}R^4\int d\tau \, ds \sin^4\theta \{ -
  \sqrt{r^2\theta^{\prime 2}+r^{\prime 2}/f(r)+(r/R)^{4}x^{\prime 2}-F_{y\theta}^2}
  +4 A_y \theta\prime \},
\ee
where $\Omega_{4}=8\pi^{2}/3$ is the volume of the unit four-sphere.

The gauge field equation of motion following from this action reads
$$
\partial_\theta D(\theta) = -4 \sin^4\theta,
$$
where the dimensionless displacement $D$ is the variation of the
action with respect to $E=F_{y\theta}$. Note, that
$F_{y\theta}$ originally was a magnetic field in the
4-dimensional theory, but became the electric field in
the three-dimensional theory, since $y$ plays now the role of time.

The solution to this equation is
\be \label{d}
D(\theta) = \left[{3\over 2}(\nu\pi-\theta)
  +{3\over 2}\sin\theta\cos\theta+\sin^{3}\theta\cos\theta\right].
\ee
As will be explained below, the integration constant $0\leq\nu\leq 1$
controls the number of Born-Infeld strings emerging from each pole of
the ${\bf S}^{5}$. Next, it is convenient to eliminate the gauge field
in favor of $D$ and Legendre transform the original Lagrangian to
obtain an energy
functional of the embedding coordinate $r(\theta)$ only:
\be \label{u}
U =  T_5 \Omega_{4}R^4\int ds
\sqrt{r^2\theta^{\prime 2}+r^{\prime 2}/f(r) +(r/R)^{4}x^{\prime 2}}\,
\sqrt{D(\theta)^2+\sin^8\theta}~.
\ee
This action has the interesting scaling property that if
$\{r(s),\theta(s),x(s)\}$ is a solution for horizon
radius $r_{h}$, then $\{\alpha r(s),\theta(s), \alpha^{-1}x(s)\}$
is a solution for horizon radius $\alpha r_{h}$. The scaling
$x\propto R^{2}/r$ is precisely as expected from the
holographic UV/IR
relation \cite{susswi,pp}. We will have more
to say about scaling behavior of solutions later on.

The fivebrane embeddings of interest to us will have singularities:
places on the five-sphere (typically $\theta\to \pi$ or $0$) where
$r\to\infty$ and $x^{\prime}\to 0$. As in~\cite{cgs,cm,gibbons},
these `spikes' must be interpreted as bundles of fundamental strings
attached to the wrapped fivebrane and localized at some definite value of $x$.
It can be seen from\req{u} that a spike sticking out at $\theta=\pi$ has a
`tension' (energy per unit radial coordinate distance)
$T_5 \Omega_{4} R^4 |D(\pi)|f(r)^{-1/2}=(1-\nu)N T_F f(r)^{-1/2}$,
which is precisely the tension of $(1-\nu)N$ fundamental strings in
this geometry. A spike at $\theta=0$ has the same tension as $\nu N$
strings so that, taken together, the two singularities
represent a total of $N$ fundamental strings, as expected. Surfaces
with more singularities and less symmetry are perfectly possible, but
a lot harder to analyze. To keep things manageable, we have built
$SO(5)$ symmetry into the ansatz.

In the extremal case ($r_{h}=0$) discussed in~\cite{cgs}, the
BPS condition provided a first integral which greatly simplified the
analysis. In the nonextremal case we are now discussing, there is
no such first integral and we have to deal with the unpleasant
second order Euler-Lagrange equation that follows from\req{u}. This
is most conveniently done in a parametric Hamiltonian formalism%
\footnote{We would like to thank G.~Savvidy for suggesting and helping
to realize this approach.}. 
Let us introduce the momenta conjugate to $r$, $\theta$ and $x$ 
\be \label{mom}
p_r=f^{-1}\dot{r}\Delta, \quad
p_{\theta}=r^2\dot{\theta}\Delta, \quad
p_{x}=(r/R)^{4}\dot{x}\Delta, \quad
\Delta=\frac{\sqrt{D^2+\sin^8\theta}}
   {\sqrt{ r^2\dot{\theta}^2 + \dot{r}^2/f+(r/R)^{4}\dot{x}^2}}~.
\ee
The Hamiltonian that follows from the action \req{u} vanishes identically
due to reparametrization invariance in $s$. Furthermore, the momentum
expressions are non-invertible and the system is subject to the constraint
\be \label{ham}
2\tilde{H} =
    \left(1-\frac{r_h^4}{r^4}\right) p_r^2 + \frac{p_{\theta}^2 }{r^2}
+\frac{R^4}{r^4}p_{x}^2-
\left( D^2+\sin^8\theta \right) =0~.
\ee
This constraint can be taken as the Hamiltonian and this choice
conveniently fixes the gauge, while getting rid of the
complicated square-root structure of the action.
The equations of motion that follow from this Hamiltonian are

\hfill
\parbox{1.7in}
{\begin{eqnarray*}
\dot r &=&\left(1-\frac{r_h^4}{r^4}\right) p_r~,\\
\dot\theta &=&\frac{p_\theta}{r^2}~, \\
\dot x &=& \frac{R^{4}}{r^{4}}p_{x},
\end{eqnarray*}}
\parbox{3.1in}
{\begin{eqnarray*}
\dot p_r &=&\frac{2}{r^5}(p_x^2 R^4 -p_r^2 r_h^4)
     + \frac{p_\theta^2}{r^3}~, \\
\dot p_\theta &=& -6\sin^4\theta \left(\pi\nu-\theta+ \sin\theta\cos\theta
\right)~,\\
\dot{p}_{x}&=& 0~.
\end{eqnarray*}}
\hfill
\parbox{0.5in}
{\bea \label{eom}
\eea}
Together with initial conditions, these equations completely define
the solutions for the fivebrane. The initial conditions should be
chosen such that $\tilde H=0$.

\begin{figure}[htb]
\centerline{\epsfxsize=14cm
\epsfbox{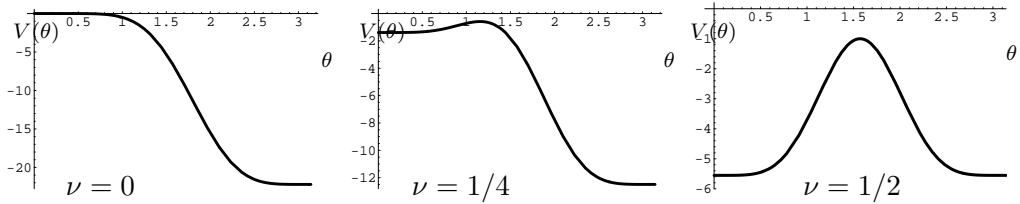}}
\caption{The potential
  $V(\theta)=-\left[D(\theta)^2+\sin^8\theta\right]$ for
  $\nu=0,1/4,1/2$ (see text for discussion).}
\label{D3pot}
\begin{picture}(0,0)
\put(12,41){\scriptsize $V(\theta)$}
\put(53,37){\scriptsize $\theta$}
\put(19,20){\small $\nu=0$}
\put(58,41){\scriptsize $V(\theta)$}
\put(98.5,37){\scriptsize $\theta$}
\put(65,20){\small $\nu=1/4$}
\put(102,41){\scriptsize $V(\theta)$}
\put(144,38){\scriptsize $\theta$}
\put(117,20){\small $\nu=1/2$}\end{picture}
\end{figure}

To gain some insight into the solutions to these equations, notice that
the basic problem to solve is a motion in the two-dimensional $r-\theta$
plane: the motion in $x$ is then determined by the choice of a conserved
value for $p_x$. Note that for $p_x=0$, the surface sits at a fixed value
of $x$ and therefore has no spatial extension in the gauge theory
coordinates: we will call this a `point' solution. When $r$ is large compared
to $r_{h}$ and $R \sqrt{p_{x}}$, 
the $(r\theta)$ motion is simply that of a particle
of unit mass moving in two dimensions under the influence of the
potential $V(\theta)=-\left[D(\theta)^2+\sin^8\theta\right]$.
By the constraint, the energy of this fictitious particle vanishes.
For generic $\nu$, the potential has three extrema (see Figure \ref{D3pot}):
two minima at $\theta=0$ and $\theta=\pi$, and a maximum at
$\theta=\theta_{c}$ such that $\pi\nu=\theta_{c}-\sin\theta_{c}\cos\theta_{c}$.
For large $r$ the particle will thus roll down towards one of the two minima.
Whether it reaches $\theta=0,\pi$ at a finite value of $r$ depends on
the initial boundary conditions.

\subsection{The Point Baryon}\label{point3dsec}

In this section we study solutions which correspond to a baryon
localized at a particular gauge theory position. To localize the 
surface
in $x$, we just set $p_{x}=0$.
With the symmetry that we have built in, the equations of motion 
typically
allow the surface to run off to $r=\infty$ at $\theta=\pi$ or $0$.
At least asymptotically, such `spikes' are equivalent to bundles of
fundamental strings and will be identified with the quark 
constituents
of the state represented by the wrapped fivebrane. To get a baryon 
whose component quarks have identical $SU(4)$ (flavor) quantum 
numbers,
we want a spike representing $N$ quarks to emerge from one pole of 
the
$S^5$ (say $\theta=\pi$) with a smooth surface at the other pole.
To meet the first condition, it suffices to set the integration 
constant
$\nu=0$ and to meet the second, we impose smooth boundary conditions
($\partial_{\theta}r=0$ and $r=r_{0}$) at $\theta=0$.\footnote{This
is equivalent to requiring $p_{r}=0$ at
$\theta=0$, in which case $p_{\theta}$ must also vanish to satisfy 
the
constraint\req{ham}.}

\begin{figure}[htb]
\centerline{\epsfxsize=12cm
\epsfbox{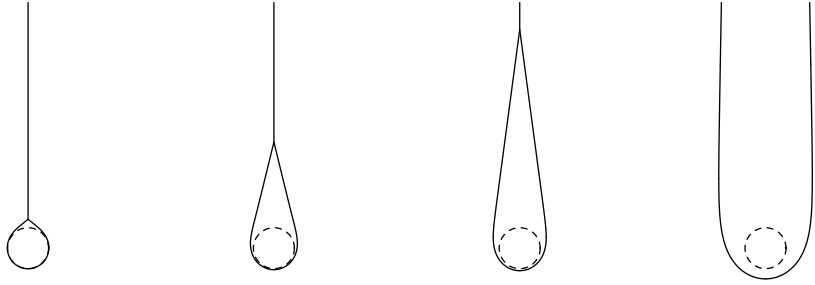}}
\caption{Family of solutions illustrating the progressive
 deformation of the fivebrane by the bundle of $N$ fundamental
 strings. The dotted  circle represents the horizon.
 The stable configuration is a `tube' with
 $r_{0}\to\infty$.}
\label{D3tubes}
\end{figure}

Numerical integration with these boundary conditions yields a
one-parameter family of solutions (parametrized by $r_0$). Due to the
scaling mentioned before, the solutions really only depend on
$r_{h}$ through the ratio $\zeta=r_{0}/r_{h}\ge 1$.
Figure \ref{D3tubes} shows polar plots\footnote{Although these plots
provide a conveniently simple representation of the solutions,
the reader should bear in mind that they are a bit misleading as to 
the
intrinsic geometry, since the radius of the ${\bf S}^{5}$ is $R$,
independent of $r$.} of the solutions for a few representative
values of $\zeta$. When $\zeta<\zeta_{crit}\approx 9$ the solution 
tilts
toward and eventually crosses the symmetry axis,
thus reaching $\theta=\pi$ at a finite value of $r$. As $\zeta\to 1$
(i.e. as the starting radius gets closer
and closer to the horizon),
the brane becomes more and more spherical (notice that
$r(\theta)=r_{h}$ is a solution to the equations of motion).
For $\zeta>\zeta_{crit}$, on the other hand,
the solution tilts away from the symmetry
axis, reaching $\theta=\pi$ only at $r=\infty$.
As $\zeta$ increases the solution looks more and more like a
`tube'. This is simply a consequence of the fact that, for large
$\zeta$, the solution is always very far from the horizon, and is
well-approximated by an extremal embedding (discussed in \cite{cgs}).
When $\zeta\to\infty$ the configuration
becomes a BPS `tube' \cite{cgs} with infinite radius.

The cusp in the $\zeta<\zeta_{crit}$ solutions indicates the presence 
of a
delta-function source in the equations of motion. Since the compactly
wrapped brane intercepts $N$ units of five-form flux, it has $N$ 
units
of worldbrane $U(1)$ charge and must have $N$ fundamental strings
attached to it \cite{wittenbaryon}. This is
most simply achieved by taking the cusp as the point of attachment of
$N$ fundamental strings, running along the ray $\theta=\pi$.
In accordance with the Born-Infeld string philosophy 
\cite{cm,gibbons},
these strings are equivalent to a D5-brane wrapped on an $S^{4}$ of 
vanishing
volume which carries the $U(1)$ flux out to infinity. A simple 
modification
of the flat space argument \cite{emp,cm,gibbons} shows that such 
a
collapsed fivebrane is a solution to the equations of motion and has
`tension' $N T_Ff(r)^{-1/2}$ (exactly the tension exerted by a
fundamental string of intrinsic tension $T_F$ in this curved space).

The entire fivebrane-string (or collapsed brane) system will be
stable only if there is tension balance between its two components.
To obtain the stability condition, let $r_{c}$ denote the location of
the cusp (which is a function of $r_{0}$), and
parametrize the family of fivebrane
embeddings as $r=r(\theta; r_{c})$. Under the variation $r_{c}\to
r_{c}+\delta r_{c}$, it can be seen from\req{u}, after an
integration by parts and application of the Euler-Lagrange equation,
that the energy of the brane changes only by a surface term,
\be \label{varcalc}
\frac{\partial U}{\partial r_{c}}=T_5 \Omega_{4}R^4
    \frac{r^{\prime}\sqrt{D^2+\sin^8\theta}}
    {f\sqrt{r^2+f(r)^{-1}r^{\prime 2}}}
    \left.\frac{\partial r}{\partial r_{c}}\right|^{\pi}_0
 =\frac{NT_Ff(r_{c})^{-1/2}}{\sqrt{1+f(r_{c}) r_{c}^2/r_{c}^{\prime
 2}}}~,
\ee
where $r_{c}^{\prime}=\partial_{\theta}r|_{\theta=\pi}$, and
we have used the fact that $r(\pi;r_{c})=r_{c}$. The numerator
in the last expression of\req{varcalc} is the `tension' at $r=r_{c}$ 
of $N$
fundamental strings, so it is clear that the brane has a
lower tension for any $r_{c}>r_{h}$. The energy is lowered by 
expanding
the fivebrane and shortening the explicit fundamental string.
A similar variational calculation applied to the $\zeta>\zeta_{crit}$
configurations (cut off at a large $r=r_{max}$) shows
that the BPS `tube' at infinity is the lowest energy solution.
This is consistent with the results of \cite{baryonsugra},
where the baryon was examined using the pure Nambu-Goto action for 
the
fivebrane wrapped on a sphere. We emphasize that the above 
variational
calculation used solutions of the full Born-Infeld (plus
Wess-Zumino) action.

Altogether, then, the solutions depicted in Fig.~\ref{D3tubes} 
provide
a physically satisfying picture of the process through which the $N$
fundamental strings deform the initially spherical fivebrane, pulling
it out to infinity. The final configuration has the shape of a 
`tube',
just like the BPS embeddings found in~\cite{imamura,cgs}. In the
supersymmetric case, $r_{0}$ was a modulus and the energy of the
baryon was independent of the overall scale of the solution.
In the nonextremal case examined here, however, there
appears to be a potential for that modulus which drives the stable 
solution
out to $r_{0}\to\infty$.

The dependence of the fivebrane embedding on the $S^{5}$ coordinates
encodes the flavor structure (i.e., the $SU(4)$ quantum numbers) of
the gauge theory state under consideration. As a result of the UV/IR
relation, the $AdS$ radial coordinate $r$ is associated with an 
energy scale
in the gauge theory, $E=r/R^{2}$ \cite{susswi,pp}. The embedding
$r(\theta)$ consequently associates a particular value of $\theta$ to
each different distance scale, yielding some sort of $SU(4)$ 
wavefunction for
the baryon. The $SO(5)$ symmetry of the embedding translates into the
statement that the baryon is a singlet under the corresponding
$SU(4)$ subgroup. Finally, the fact that a given surface spans
the range $r\ge r_{0}$ implies that the dual gauge theory 
configuration
has structure on all length scales from zero up to a characteristic 
size
$R^{2}/r_{0}$. Since the energetically preferred configuration has 
$r_{0}\to\infty$, it is in this sense truly pointlike.

\subsection{The Split Baryon: Color Dependence of the String Tension}
\label{split3dsec}
We now turn our attention to solutions with $p_{x}\neq 0$
(i.e. $x^{\prime}\neq 0$). They describe collections of quarks
at finite separation in the gauge theory position space and are of
interest for exploring confinement issues. It turns out to
be rather easy to construct a surface describing an $SU(N)$
baryon split into two distinct groups, containing $\nu N$ and
$(1-\nu)N$ quarks respectively and separated by a distance $L$
in the $x$ direction. In a
confining $SU(N)$ gauge theory, two such quark bundles should be
connected by a color flux tube and we will study the Born-Infeld
representation of this phenomenon. Each group of quarks corresponds
as before to a bundle of Born-Infeld strings, realized in our
approach as a singular spike or fivebrane `tube' with topology
${\bf R}\times {\bf S}^{4}$. Remember that we have assumed an
$SO(5)$-symmetric configuration, which means that the two
singularities representing the quarks must be located at opposite
poles of the ${\bf S}^{5}$ (we will put them at $\theta=0$ and
$\theta=\pi$) with corresponding implications about the $SU(4)$
flavor structure of the states we are constructing. More general
flavor structures are possible, but we will not try to study these
more complicated surfaces. For large spatial separation
$L=\vert x(\infty)-x(-\infty)\vert$, the portion of the fivebrane 
that
interpolates between the two string bundles runs close to the
horizon and it is this part of the surface that encodes the
confining flux tube of the gauge theory. The surface equations 
\req{eom}
imply that the part of the surface that has large spatial extent
must sit at a constant $\theta=\theta_{c}$ where $\dot p_{\theta}=0$.
More precisely, it has to sit at the solution of
\be \label{nu}
\pi\nu=\theta_{c}-\sin\theta_{c}\cos\theta_{c}~.
\ee
corresponding to the unstable maximum of the
potential $V(\theta)$ discussed at the end of Section \ref{eqn3dsec}.
The critical angle is a monotonic function of $\nu$, such that
$\theta_{c}(0) =0$ and $\theta_{c}(\nu)=\pi-\theta_{c}(1-\nu)$.
The energetics of the part of the fivebrane that encodes the 
confining
flux tube will depend on $\theta_c$, and therefore $\nu$, in a way 
that
we will now examine in some detail.

\begin{figure}[htb]
\centerline{\epsfxsize=6cm
\epsfbox{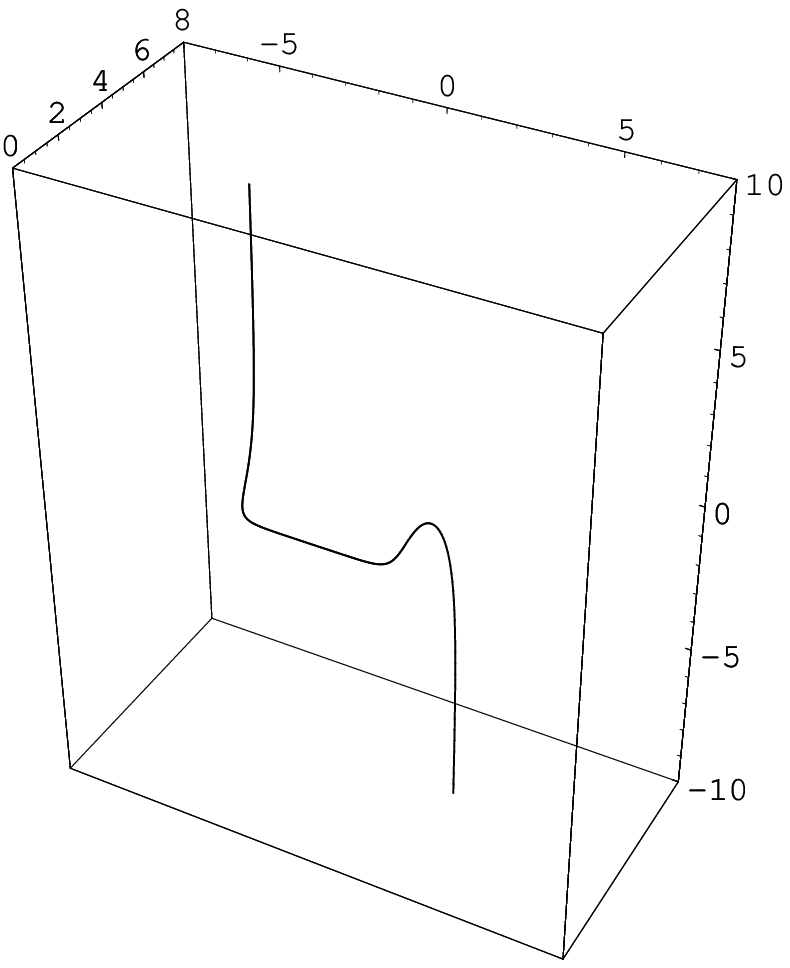}}
\begin{picture}(0,0)
\put(35,72){\small $r\sin\theta$}
\put(85,75){\small $x$}
\put(106,39){\small $-r\cos\theta$}
\end{picture}
\caption{The three-dimensional projection of the D5-brane.
Every point on the curve is an ${\bf S}^{4}$.
One can clearly see how the brane drops down towards the horizon,
extends horizontally along it, and finally leaves at the other end.
{}From the point of view of the three-dimensional $SU(N)$ gauge 
theory,
which lives in the $\{x,y,z\}$ directions, this configuration
represents a baryon split into two groups of
$\nu N$ and $(1-\nu)N$ quarks (the vertical segments
--- see Fig.~\ref{splitpol}),
connected by a flux tube extending a finite distance along the $x$
direction.}
\label{split3d}
\end{figure}

\begin{figure}[htb]
\centerline{\epsfxsize=6cm
\epsfbox{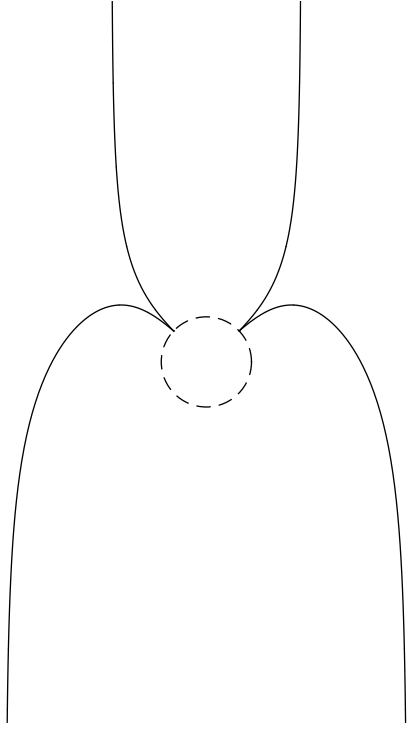}}
\caption{Polar plot of $r(\theta)$ -- the two-dimensional
projection of Fig.~\ref{split3d}.
The Born-Infeld string
`tubes' pointing up and down represent two groups of $\nu N$ and 
$(1-\nu)N$
quarks, respectively (with $\nu=0.9$ here).
The brane extends in the $x$-direction mostly
at the inflection point (the cusp), while
$x$ is essentially constant along the tubes.}
\label{splitpol}
\end{figure}

Unfortunately, we must resort to numerical analysis to construct 
specific
surfaces of this kind. It is convenient to take the point of closest 
approach to the horizon as the starting point for the numerical 
integration. 
We start the integration off with the initial conditions
\be \label{inicon}
\begin{array}{ll}
r(0)= r_h + \epsilon~,&  p_r(0)=0~,\\
\theta(0)=\theta_{c}~,&  p_{\theta}(0)=\eta~,\\
x(0)=0~,&  p_x(0)=(r(0)/R)^{2}\sqrt{\sin^6\theta_{c}
-(\eta/ r(0))^2}~.
\end{array}
\ee
The distance from the horizon at the point of closest approach is 
controlled
by $\epsilon$. For a given $\epsilon$, we have to `shoot' in $\eta$ 
until
we get
satisfactory behavior of the quark-like singularities at
$\theta\rightarrow 0$ and $\theta \rightarrow \pi$ (see Section
\ref{point3dsec} for details). Indeed, it is natural to require
asymptotically BPS behavior in the region of space
where supersymmetry is recovered locally (i.e., far from the 
horizon).
Once that is done, $\epsilon$ controls the spatial separation of the 
two
separated quark bundles. Figs.~\ref{split3d} and \ref{splitpol}
depict a typical fivebrane embedding obtained by numerical 
integration,
for the case $\nu=0.9$~. It can be seen in Fig.~\ref{split3d} that 
the brane
extends in the $x$ direction mostly in its `flux tube' portion, at
$\theta=\theta_{c}$ and $r\approx r_{h}$. The Born-Infeld string 
`tubes'
corresponding to the two groups of quarks lie essentially at a 
constant
value of $x$. 

It is seen from the equation for $\dot r$ in (\ref{eom}) that, for 
the
portion of the brane running close to the horizon, $r(s)-r_h$
grows as an exponential in the parameter $s$,
with an exponent proportional to $p_r$. The latter
reaches a value close to $(R/r_{h})p_x$ (see the equation for $\dot 
p_r$).
The equation for $\dot x$ then shows that the separation $L$
between the quarks increases only logarithmically with $\epsilon$, 
the
minimal distance to the horizon. In fact, there
exists a limiting solution which consists of $\nu N$ quarks with a 
flux tube
that extends to infinity, and the brane approaches the horizon 
exponentially
with distance.

{}From the above discussion it is clear that for large quark 
separation $L$
the (renormalized) energy will receive its main contribution from the 
flux
tube, and will consequently depend linearly on $L$, a clear 
indication of 
confinement.  It is easy to compute the tension (energy per unit 
distance in $x$) of the color flux tube.
Note that when the fivebrane runs parallel
and close to the horizon, the energy function (\ref{u}) reduces
to
$$U_{flux}=\frac{2}{3\pi} N T_{F}
   \int dx \left(\frac{r}{R}\right)^{2}
    \sqrt{D^2 + \sin^8 \theta}~. $$
Using $r\simeq r_{h}$, $\theta\simeq\theta_{c}$, and performing some 
simple
manipulations using the definitions of $\theta_c$\req{nu} and 
$D(\theta)$\req{d}, one obtains the tension
\be \label{tension}
\sigma_{3}(\nu)=\frac{2}{3\pi}N T_{F} \left(\frac{r_h}{R}\right)^{2}
  \sin^{3}\theta_{c}=\frac{\sqrt{2}}{3}N\sqrt{\g{3}^{2}N}T^{3/2}
  \sin^{3}\theta_{c}~,
\ee
where the last expression is given solely in terms of parameters of
the gauge theory in three dimensions. Since it is obtained by
dimensional reduction from four dimensions, this theory should be
understood to have an ultraviolet cutoff proportional to
the Hawking temperature $T$.
The dependence of the tension on $T$, $N$, and the 't~Hooft
coupling $\lambda_{3}=\g{3}^{2}N$ agrees with the result
of~\cite{baryonsugra}, where a baryon whose component quarks lie on
a circle is treated within a simplified
Nambu-Goto approach.

Notice that, in addition, equation\req{tension}
gives the dependence of the flux tube tension
on $\nu$, i.e. on its color content. This nontrivial 
dependence, arising entirely from the factor
$\sin^{3}\theta_{c}$, is plotted in Fig.~\ref{D3tension}:

\vspace{0.4cm}

\begin{figure}[htb]
\centerline{\epsfxsize=8cm
\epsfbox{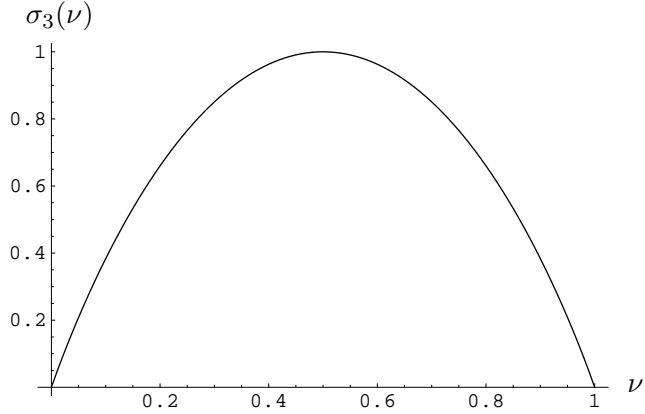}}
\begin{picture}(0,0)
\put(40,57){\small $\sigma_3(\nu)$}
\put(120,8){\small $\nu$}
\end{picture}
\caption{The tension of the flux tube (normalized to unity
 at its peak) as a function of
 $\nu$, the fraction of quarks pulled apart.
 In the full theory $\nu$ should be quantized in units of ${1/N}$.
  See text for discussion.}
\label{D3tension}
\end{figure}

Let us discuss the features of the flux tube
tension seen in Fig.~\ref{D3tension}.
As one might expect, the tension increases linearly for small $\nu$.
This means that for each additional quark removed the work done is
approximately constant.
Also as expected, in expression\req{tension} there is a
complete symmetry between $\nu$ and $1-\nu$, (i.e., it makes no 
difference
whether $n$ or $N-n$ quarks are pulled out). Thus, the
tension has a maximum
at $\nu=1/2$ and comes back down to zero near $\nu=1$. In gauge 
theory
language,
the flat part of the curve means that it does not cost any energy to 
move
the quark from one bundle with $\sim N/2$ quarks to the other.

Notice from equation\req{nu} that $\sin^{3}\theta_{c}\simeq 
3\pi\nu/2$
for small $\nu$. This implies that $\sigma_{3}(\nu=1/N)$ becomes {\em
independent} of $N$ in the 't~Hooft limit $N\to\infty$ with
$\lambda_{3}$ fixed.  This result has a natural
gauge theory interpretation.  When one quark is pulled out from the
$SU(N)$ baryon (a color-singlet), the remaining $N-1$ must be in the
anti-fundamental representation of the gauge group.  The flux tube
extending between this bundle and the solitary quark should then have
the same properties as the standard QCD string which connects a quark
and an antiquark.  In particular, its tension should depend on $N$
only through the 't~Hooft coupling, as we have found. As a matter of
fact, for $\nu=1/N\ll 1$ equation\req{tension} precisely agrees with
the quark-antiquark string tension which follows from a Nambu-Goto
string calculation \cite{bisy2}.  More generally, for $\nu=n/N$, with
$n$ fixed as $N\to\infty$, expression\req{tension} reduces to the
tension of $n$ quark-antiquark strings.

It is important to note that, as has been pointed out by various 
authors,
the gauge theory under study here is not strictly
three-dimensional \cite{bisy2, groguri}.
The energy scale associated with the QCD string
tension, for instance, is proportional to $\lambda_{4}^{1/4}T$, where 
$\lambda_{4}=\g{4}^{2}N$.
This is much larger than the
compactification scale $T$ in the large $\lambda_{4}$ regime where
the supergravity approximation is appropriate.

\newpage
\addcontentsline{toc}{chapter}{Conclusion}

\addtolength{\baselineskip}{-.1cm}
\begin{center}
{\huge\bf Conclusion}
\end{center}

In conclusion I would like to sum up what has been done in this thesis,
and point out some interesting possible extensions of our work.

In the second Chapter I conclusively show the broad applicability
of the Born-Infeld action to static and dynamical questions in string theory.

In the third Chapter I present the work done in collaboration with
C. Callan, A. Guijosa, and \O. Tafjord on application of the
B-I fivebrane worldvolume action to the AdS/CFT correspondence.
We construct a non-perturbative
object of gauge theory, namely the baryon vertex. 

Our description of the Hanany-Witten phenomenon raises a very
interesting question which I would like to explore in the future,
whether the half-string state that can be conveniently described by
our B-I wrapped brane has in fact reality to it, i.e. 
what are the dynamical consequences of such an object,
and its place in perturbative string theory.

In the second part of the Chapter I present our work on the
supergravity/gauge theory correspondence in the case of
non-extremal background. Here we were able to measure 
the tension of the color flux tubes, and its dependence
on the number of quarks removed from the baryon to a 
finite distance. 

The possible extension of these efforts 
to get a handle at weakly coupled non-supersymmetric theory
would be to the new ideas coming from type 0A and 0B theories 
\cite{klebanov,minahan,others}.
In these theories spacetime fermions are absent from the very
beginning and there is no need to invent ways to decouple them.
Moreover there are solutions to the gravity equations which
have a running dilaton, and on the gauge theory side the 
running coupling matches exactly the expectation from ordinary QCD.

It is at present too early to comment on the validity 
of this approach, for example because of the problems associated
with the presence of the tachyon in the low energy spectrum.


\addcontentsline{toc}{chapter}{Acknowledgements}
\vfill
{\large\bf Acknowledgements}
\vspace{2cm}
Before everyone else my Parents, Natalia and George, 
and my brother Pavlos have given me
the most valuable support and encouragement whenever it was most needed.
Also, the people of my extended family have been a model of what a big heart,
golden hands and hard work can achieve. It is to that simple model
that I strive to measure up to.

I would like to thank everybody at Princeton who I worked with,
was helped by, or just socialized. I have changed much and
learned much over these years both scientifically and otherwise.
The people who are most responsible for this, and to whom I am
most thankful are my advisor Curtis Callan, as well as Igor Klebanov
and Larus Thorlacius.
Earlier I  benefited tremendously  from discussions with
David Gross, Washington Taylor, and Matt Strassler.

Also, many thanks to the students in our department, discussions
with whom played a big role in learning things: my coworkers
Alberto Guijosa and Oyvind Tafjord; John Brodie, Andrei
Mikhailov, Shiraz Minwalla, Sangmin Lee, Vamsi Madhav,
Maulik Parikh, Morten Krogh, Aki Hashimoto,
Steve Gubser, Sergei Gukov, Victor Gurarie, Michael Krasnitz,
Juan Maldacena, Matt Hastings, Edna Cheung and several others.

I would also like to thank my friends who are too numerous to
list all, Dimitris Kyritsis, Alexandros Michaelides, Angelos Bilas,
Andromache Karanika, Diego Casa, Sanjay Ram, Tina Pavlin,
Jacqui Hall, Marianna Shlyakhova, Paul Shultz, Jamie Fumo, 
Areg and Shooshan Danagoulian, Dima Oulianov, 
Cristina and Jorge Silva, Sasha Shekochikhin, Florin Spinu,
Bogdan Ion, Evi, Anthi, George, Alexis,
Max Perelstein, Billy Bridges, Elcin Yildirim, Partha Nandi,
Don Priour, Antonio Garcia, Davee Datta, Stas Boldyrev,
Anagha Neelakantan and many others.

\end{document}